\documentclass{article}

\usepackage{amsmath}
\usepackage{graphicx}
\usepackage{natbib}
\usepackage{siunitx}
\DeclareSIUnit\hPa{\hecto\pascal}
\DeclareSIUnit\year{yr}

\title{State space models for non-stationary intermittently coupled systems: an application to the North Atlantic Oscillation}
\author{Philip G. Sansom, Daniel B. Williamson, David B. Stephenson}

\newcommand{\bB}{\mathbf{B}}              
\newcommand{\bh}{\mathbf{h}}              
\newcommand{\bH}{\mathbf{H}}              
\newcommand{\bm}{\mathbf{m}}              
\newcommand{\bC}{\mathbf{C}}              
\newcommand{\ba}{\mathbf{a}}              
\newcommand{\bR}{\mathbf{R}}              
\newcommand{\bA}{\mathbf{A}}              
\newcommand{\bF}{\mathbf{F}}              
\newcommand{\bG}{\mathbf{G}}              
\newcommand{\bw}{\mathbf{w}}              
\newcommand{\bW}{\mathbf{W}}              
\newcommand{\bY}{\mathbf{Y}}              
\newcommand{\bJ}{\mathbf{J}}              

\newcommand{\bzero}{\mathbf{0}} 

\newcommand{\btheta  }{\boldsymbol{\theta  }}  
\newcommand{\bphi    }{\boldsymbol{\phi    }}  
\newcommand{\bmu     }{\boldsymbol{\mu     }}  
\newcommand{\bSigma  }{\boldsymbol{\Sigma  }}  
\newcommand{\taub    }{\boldsymbol{\taub   }}  
\newcommand{\bPhi  }{\boldsymbol{\Phi  }}  

\DeclareMathOperator{\Tri}{\mathit{Tri}}   
\DeclareMathOperator{\B}{\mathit{Beta}}   
\DeclareMathOperator{\N}{\mathit{N}}      

\newcommand{\Bp}[2]{\B \left( #1, #2 \right)}           
\newcommand{\Np}[2]{\N \left( #1, #2 \right)}           

\newcommand{\Prp}[1]{\Pr \left( #1 \right)}               

\DeclareMathOperator{\E}{E}              

\newcommand{\Ep}[1]{\E \left( #1 \right)}                 

\begin{document}

\maketitle

\begin{abstract}
We develop Bayesian state space methods for modelling changes to the mean level or temporal correlation structure of an observed time series due to intermittent coupling with an unobserved process.
Novel intervention methods are proposed to model the effect of repeated coupling as a single dynamic process.
Latent time-varying autoregressive components are developed to model changes in the temporal correlation structure.
Efficient filtering and smoothing methods are derived for the resulting class of models.
We propose methods for quantifying the component of variance attributable to an unobserved process, the effect during individual coupling events, and the potential for skilful forecasts.

The proposed methodology is applied to the study of winter-time variability in the dominant pattern of climate variation in the northern hemisphere, the North Atlantic Oscillation.
Around \SI{70}{\percent} of the inter-annual variance in the winter (Dec-Jan-Feb) mean level is attributable to an unobserved process.
Skilful forecasts for winter (Dec-Jan-Feb) mean are possible from the beginning of December.
\end{abstract}

\section{Introduction}
\label{sec:introduction}

Intermittently coupled systems can be found in many areas of both the natural and social sciences.
We define an intermittently coupled system as one which can be modelled by two or more component processes which only interact at certain times.
For example, many climate processes are only active during certain times of year, e.g., sea ice and snow cover change the interaction between the surface and the atmosphere \citep{Chapin2010,Bourassa2013}.
Migrating birds and animals only mix at certain times of year, allowing disease transmission between populations \citep{Olsen2006,Altizer2011}.
Empirical models have been applied to forecasting intermittent demand in production economics and operational research \citep{Croston1972,Shenstone2005}.
Interest will often focus on one component of the system, while the others may be impossible or impractical to observe or even to physically identify. 
However, physical reasoning or prior knowledge may support the existence of such components, and provide information about their behaviour and their effect on the component of interest.
We refer to these secondary processes as intermittently coupled components, and the times at which the processes interact as coupling events.
By incorporating this information through careful statistical modelling we can separate the effect of intermittently coupled components from the underlying behaviour of the observed system.   

The methodology developed in this study was motivated by the problem of diagnosing unusual persistence in the dominant mode of climate variability in the northern hemisphere, known as the North Atlantic Oscillation (NAO).
Because of its impact on European climate, the ability to forecast the NAO is currently a topic of great interest for the development of new climate prediction services \citep{Siegert2016}.  
A daily time series of NAO observations is shown in Fig.~1.
There is a clear annual cycle in the observations.
Figure~1 also indicates that the NAO exhibits greater inter-annual variability in the extended winter season (Dec--Mar) than the extended summer season (Apr--Nov).
At the same time, the autocorrelation function indicates increased persistence of day-to-day conditions between December and March than between April and November.
Increased persistence, implies increased predictability.
The seasonal contrast in inter-annual variability and autocorrelation visible in Fig.~1 could be caused by a transient shift in the mean, or a change in autocorrelation structure during between December and March.
Climate scientists typically fit separate models to different seasons \citep[e.g.,][]{Keeley2009,Franzke2011}.
This approach makes it difficult to diagnose whether the apparent change in autocorrelation is the cause of the increased inter-annual variability, or a symptom of it.

\begin{figure}[t]
  \includegraphics[width=\textwidth]{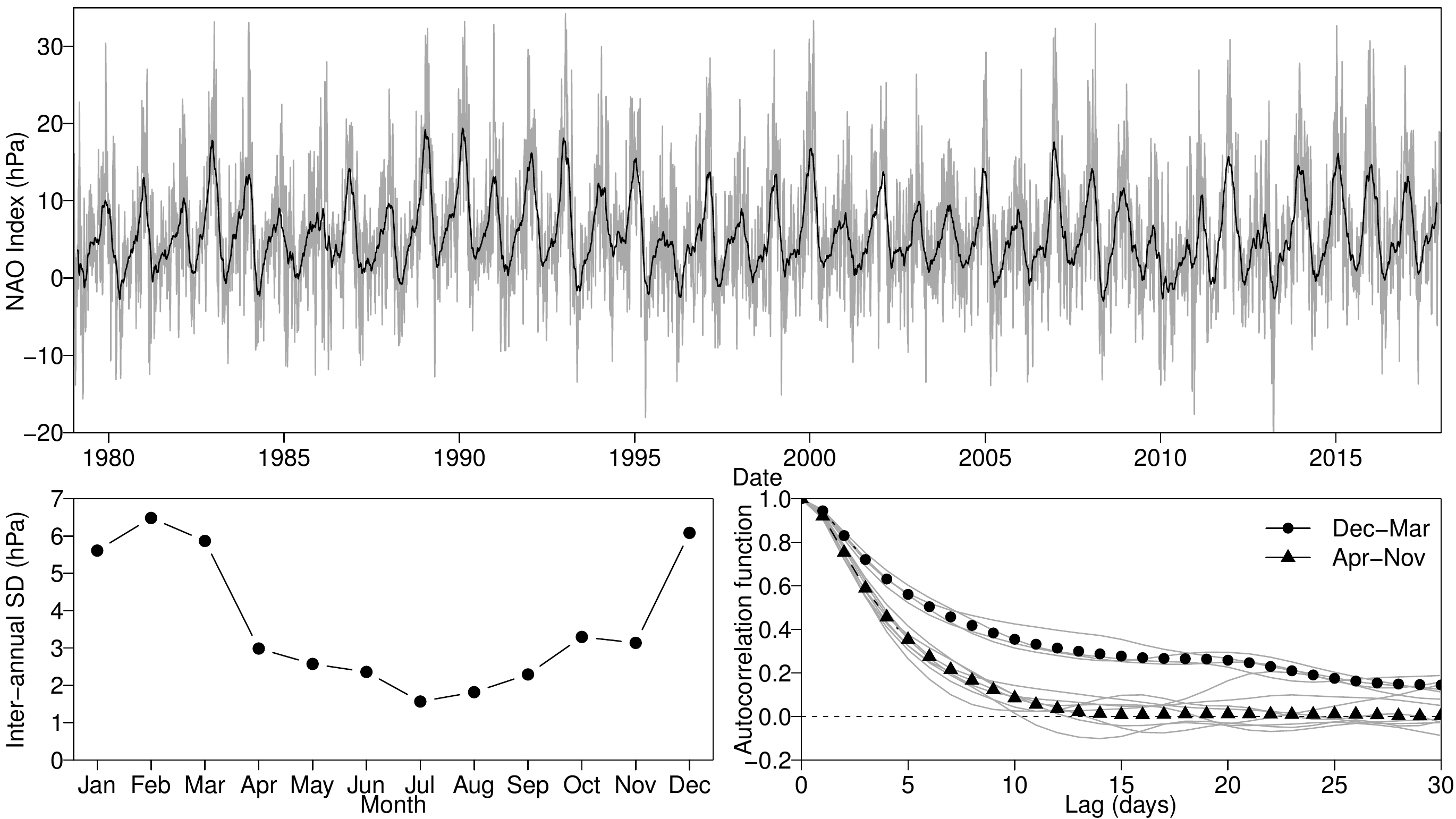}
  \caption{The North Atlantic Oscillation. 
           (top) Daily time series (grey) and 90 day moving average (black) of our daily NAO index, 
           (left) inter-annual standard deviation of the monthly mean NAO index, and
           (right) the autocorrelation function of the daily NAO index computed for each month of the year.
           Grey lines are the individual months.
           A linear trend and annual and semi-annual cycles were estimated by least-squares and removed before computing the autocorrelation functions.
           \label{fig:nao}}
\end{figure}

In this study we propose a flexible class of models capable of separating variability due to unobservable intermittent components, from long term variability in the observed process itself, accumulated short-term variability, and observation errors.
We develop tools for diagnosing whether the intermittent component acts on the mean or the autocorrelation structure of the observed system.
If we can learn the state of the intermittent component quickly enough, then it should be possible to make skilful predictions about the remainder of a particular coupled period.
Alternatively, the effect of the intermittent component may be similar between coupling events.
In that case, it should be possible to make predictions about subsequent coupled periods.

State space models, also known as structural time series models, provide a flexible class of models for non-stationary time series \citep{DurbinKoopman}.
By modelling the system in terms of physically meaningful components we can incorporate expert knowledge to help separate the effects of intermittent components from long-term variability elsewhere in the system.
There is an extensive literature on modelling non-stationarity in the mean by state space methods, particularly where the observed process depends linearly on the state parameters and the observation and evolution processes are both normally distributed \citep{Harvey,WestHarrison,DurbinKoopman}.
Time-varying autoregressive (TVAR) models generalise classical autoregressive models to have time-varying coefficients, thus capturing changes in the autocorrelation structure \citep{Rao1970,Kitagawa1985,Prado1997,PradoWest}.
In Sec.~\ref{sec:tvar}, we propose a class of models containing latent TVAR components that capture changes in short-term temporal dependence while maintaining the interpretability of the mean and unobserved intermittent effects.

Smooth changes in the mean or the temporal dependence structure can be captured by simple random walk priors on their respective state variables.
Rapid changes, such as those that might be expected due to intermittent coupling, often require explicit interventions in the model \citep{Box1975}.
Intervention methods were extended to state space models by \citet{Harvey1986}.
Standard intervention approaches (e.g., \citet[Chapter 7.6]{Harvey}, \citet[Chapter 11]{WestHarrison}, \citet[Chapter 3.2.5]{DurbinKoopman}) require the introduction of separate intervention and effect variables for each event.
The effect is usually assumed to be constant throughout a particular event and independent between events.
In the case of intermittent coupling, the underlying cause of each event will usually be the same, although the effect may vary.
In Sec.~\ref{sec:coupling}, we model the effect of intermittent coupling as a single dynamic process, intermittently identifiable through a series of interventions that determine the timing and duration of the coupling events.

The construction of the NAO time series shown in Fig.~1 and analysed in Sec.~\ref{sec:results} is described in Sec.~\ref{sec:nao}.
Following the methodological developments outlined above, we discuss efficient posterior inference for the resulting class of models in Sec.~\ref{sec:inference}.
Section~\ref{sec:results} contains the results of our study of the NAO.
Section~\ref{sec:discussion} concludes with a discussion.

\section{The North Atlantic Oscillation}
\label{sec:nao}

The North Atlantic Oscillation is the name given to the difference in surface pressure between the Azores High and the Icelandic Low \citep{Walker1924}.
The NAO is important because it affects the strength of the prevailing westerly winds and the position of the storm track, strongly influencing the winter climate of the United Kingdom and Europe \citep{Hurrell1995}.
The NAO varies on time scales from a few days to several decades \citep{Hurrell1995,Kushnir2006}.
Statistical studies have hinted at the potential to predict the NAO on seasonal time scales \citep{Keeley2009,Franzke2011}.
This potential predictability is often attributed to forcing by slowly varying components of the climate system, including sea surface temperatures, the stratosphere and snow cover \citep{Kushnir2006}.
Climate models have recently begun to show significant skill in forecasting the winter NAO a season ahead \citep{Scaife2014}.
However, the physical mechanisms behind the predictability remain unclear and the size of the predictable signal appears to be underestimated by the models \citep{Scaife2014,Eade2014}.
Careful statistical modelling may lead to additional insights.
If a predictable signal can be extracted from the observations, then it may be possible to identify the source of the forcing effect.

Following \citet{Mosedale2006}, we construct a simple NAO index as the area-weighted sea level pressure difference between two boxes, one stretching from $20^\circ$--$55^\circ$N, the other from $50^\circ$--$90^\circ$N, both spanning $90^\circ$W--$60^\circ$E, using pressure data from the ERA-Interim reanalysis \citep{Dee2011}.
The resulting daily time series, shown in Fig.~\ref{fig:nao}, spans the period 1 January 1979 to 31 December 2017, a total of $T=14\,245$ observations.

\section{Modelling intermittently coupled systems}
\label{sec:model}

In complex systems such as the Earth system, it is reasonable to consider that all components of the system (e.g., mean, seasonality, temporal dependence) may evolve slowly over time.
We begin by outlining a general model to capture gradual changes in the underlying components of the observed process.
We then propose explicit intervention models to represent rapid transient changes due to intermittent coupling.

\subsection{Latent TVAR component models}
\label{sec:tvar}

Classical autoregressive models require that we redefine the mean of the observed process, if the mean is non-zero. 
This makes it difficult to specify physically meaningful models for the time evolution of the mean and the effect of intermittently coupled components.
Latent autoregressive components remove the need to redefine the mean level of the observed time series \citep[Chapter~2]{Harvey}.
In order to allow for possible changes in the mean, seasonal and autocorrelation structure of an observed process, we propose the following latent \emph{time-varying} autoregressive component model with observation equation
\begin{align}
  Y_t & =    \mu_t + \textstyle\sum_{k = 1}^K \psi_{k t} + X_t + v_t &
  v_t & \sim \Np{0}{V} & k & = 1,\ldots,K
  \label{eqn:observation}
\end{align}
and evolution equations
\begin{subequations}
  \begin{align}
    \mu_t            & = \mu_{t-1} + \beta_t + w_{\mu t}  &
      w_{\mu t}          & \sim \Np{0}{W_\mu} \\ 
    \beta_t          & = \beta_{t-1} + w_{\beta t}  &
      w_{\beta t}        & \sim \Np{0}{W_\beta} \\
    \psi_{k t}       & = \psi_{k,t-1}       \cos k \omega 
                       + \psi_{k,t-1}^\star \sin k \omega
                       + w_{\psi_k t} &
      w_{\psi_k t}       & \sim \Np{0}{W_\psi} & k & = 1,\ldots,K \\
    \psi_{k t}^\star & = \psi_{k,t-1}^\star \cos k \omega   
                       - \psi_{k,t-1}       \sin k \omega
                       + w_{\psi_k^\star t} &
      w_{\psi_k^\star t} & \sim \Np{0}{W_\psi} & k & = 1,\ldots,K \\
    X_t              & = \textstyle\sum_{p=1}^P \phi_{p t} X_{t-p} + w_{X t} &
      w_{X t}            & \sim \Np{0}{W_{X t}}  \\
    \phi_{p t}       & = \phi_{p,t-1} + w_{\phi_p t} &
      w_{\phi_p t}       & \sim \Np{0}{W_\phi} & p &= 1,\ldots,P
  \end{align}
  \label{eqn:evolution}
\end{subequations}
where $\omega = 2 \pi / 365.25$.
The observed process $Y_t$ is modelled as the sum of mean, seasonal and autoregressive components.
The variable $\mu_t$ represents the mean level of the observed process.
Any local-in-time systematic trend is captured by the variable $\beta_t$.
The harmonic components $\psi_{kt}$ and $\psi^\star_{kt}$ ($k = 1,\ldots,K$) represent seasonal behaviour.
The local trend and seasonal variables are assumed to be time-varying, evolving according to independent normal evolution processes $w_{\mu t}$, $w_{\beta t}$, $w_{\psi_k t}$ and $w_{\psi_k^\star t}$ ($k = 1,\ldots,K$). 
The irregular component $X_t$ represents short-term variability in the observed process and is modelled as a latent time-varying autoregressive process of order $P$ with normal evolution process $w_{X t}$.
The autoregressive coefficients $\phi_{p t}$ are assumed to be time-varying, evolving according to independent normal evolution processes $w_{\phi_p t}$ ($p = 1,\ldots,P$).
The independent residual $v_t$ represents observation or measurement error.

In the case of the NAO, the variance $W_{X t}$ of the evolution process $w_{X t}$ is expected to vary systematically with the solar cycle and is modelled as 
\begin{align}
  W_{X t} & = W_X + \sqrt{a^2 + b^2}+ a \sin \omega t + b \cos \omega t &
    W_X > 0.
  \label{eqn:wxt}
\end{align}
The other evolution and error variances $W_\mu$, $W_\beta$, $W_\psi$, $W_\phi$ and $V$ are assumed constant over time.
Model (\ref{eqn:evolution}) is intended to capture gradual changes in the structure of the observed process.
Therefore, the evolution variances $W_\mu, W_\beta, W_\psi$ and $W_\phi$  are expected to be small, in particular $W_\mu,W_\beta,W_\psi,W_\phi \ll W_{X_t}$.
The evolution and error variances are assumed unknown and must be inferred from the data.
The variance parameters $W_x$, $a$ and $b$ of the irregular component in (\ref{eqn:wxt}) must also be inferred from the data.
Expert judgement about the scale of the evolution variances can incorporated through appropriate prior probability distributions.

The model defined by (\ref{eqn:observation}) and (\ref{eqn:evolution}) is quite general and could be applied to a range of climate, economic or environmental time series.
Examination of the sample periodogram for our NAO index showed clear evidence of  six and 12 month cycles, suggesting a model with $K = 2$ harmonic components.
Examination of the sample autocorrelation and partial autocorrelation functions suggest a latent TVAR process with $P = 5$ coefficients (after removing a linear trend, and constant annual and semi-annual cycles estimated by least squares).

\subsection{Intervention methods for intermittent coupling}
\label{sec:coupling}

\begin{figure}[t]
  \centering
  \includegraphics[width=0.5\textwidth]{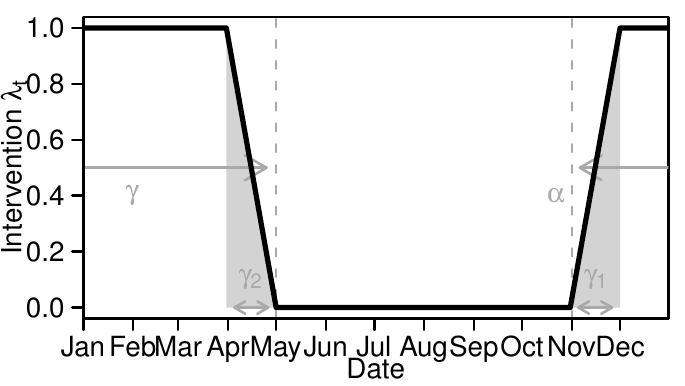}
  \caption{Example of the form and parametrisation of the intervention $\lambda_t$. 
           Parameters $\alpha$ and $\gamma$ represent the start and duration of the coupled period, while $\rho = (\gamma_1 + \gamma_2) / \gamma$ controls the transition.
           \label{fig:intervention}}
\end{figure}

The change in the autocorrelation structure of the NAO index in Fig.~\ref{fig:nao} appears to involve two distinct states, i.e., coupled or not.
We model the change from the uncoupled to the coupled state by introducing an intervention variable
\begin{equation*}
  \lambda_t = 
    \begin{cases}
      0 & \text{if} \ t \notin \boldsymbol{\tau} \\
      1 & \text{if} \ t \in    \boldsymbol{\tau}
    \end{cases}
\end{equation*}
where $\boldsymbol{\tau}$ is the set of times $t$ where the observed system is believed to be coupled to the unobserved process, e.g., $\boldsymbol{\tau} = \lbrace \text{Dec,Jan,Feb,Mar} \rbrace$.
We assume that the timing and duration of the coupling events is constant between events, but not known precisely.
We model the intervention $\lambda_t$ by introducing two hyper-parameters $\alpha$ and $\gamma$ representing the start and duration of the coupled period $\boldsymbol{\tau}$ respectively (Fig.~\ref{fig:intervention}).
In practice, we do not expect an instantaneous change in the behaviour of the system.
Therefore, we linearly taper the intervention $\lambda_t$ from zero to one over a period $\gamma_1$ at the start of the coupled period and from one to zero over a period $\gamma_2$ at the end of the coupled period.
In the absence of stronger beliefs, we assume the tapering is symmetric (i.e., $\gamma_1 = \gamma_2$) and accounts for a proportion $\rho = (\gamma_1 + \gamma_2)/\gamma$ of the duration $\gamma$.
The hyper-parameters $\alpha$, $\gamma$ and $\rho$ are assumed to be unknown and must be inferred from the data.

We consider two alternative models for the effect of intermittent coupling.
First, coupling may lead to a transient change in the mean of the observed process, second, coupling may lead to a transient change in the temporal dependence structure of the observed process.
If coupling is believed to induce a change in the mean, then the forecast equation (\ref{eqn:observation}) is modified to include the intervention as follows
\begin{align}
  Y_t & = \mu_t + \textstyle\sum_{k = 1}^K \psi_{k t} + X_t + 
          \lambda_t \delta_t + v_t &
    v_t & \sim \Np{0}{V}.
  \label{eqn:fdelta}
\end{align}
The effect $\delta_t$ is modelled as
\begin{align}
  \delta_t & = \varphi \delta_{t-1} + w_{\delta t} &
    w_{\delta t} & \sim \Np{0}{W_\delta}.
  \label{eqn:delta}
\end{align}
We interpret the effect $\delta_t$ as a change in the mean level, since we expect the effect variance to be small, i.e., $W_\delta \ll W_{X_t}$.
However, it should be noted that when $\lambda > 0$, the day-to-day variability of the observed process $Y_t$ will increase slightly, in addition to any systematic change captured by $W_{X_t}$ in (2).

If coupling is believed to induce a change in the autocorrelation structure, then we modify the forecast equation (\ref{eqn:observation}) again
\begin{align}
  Y_t & = \mu_t + \textstyle\sum_{k = 1}^K \psi_{k t} + X_t + 
          \lambda_t \textstyle\sum_{p=1}^P \delta_{p t} X_{t-p} + v_t &
    v_t & \sim \Np{0}{V}
  \label{eqn:fdeltap}
\end{align}
and define $P$ effects $\delta_{p t}$, modelled as 
\begin{align}
  \delta_{p t} & = \varphi \delta_{p,t-1} + w_{\delta_p t} &
    w_{\delta_p t} & \sim \Np{0}{W_\delta} &
    p & = 1,\ldots,P
  \label{eqn:deltap}
\end{align}
with common hyper-parameters $\varphi$ and $W_\delta$.

Most of our prior knowledge about coupling events is likely to be about their timing, and will be expressed through priors on the hyper-parameters $\alpha$, $\gamma$ and $\rho$.
Therefore, it is difficult to justify a complex form for the effects $\delta_t$ or $\delta_{p t}$.
However, a variety of behaviour can be captured depending on the values of the coefficient $\varphi$ and variance $W_\delta$.

As noted in the previous section, the mean, trend, seasonal and autoregressive parameters are expected to vary only slowly.
Therefore, we can learn their states outside of the coupled period and identify the coupling effects $\delta_t$ or $\delta_{p t}$ ($p=1,\ldots,P$) when $\lambda_t > 0$.
The form and parametrisation of the coupling intervention $\lambda_t$ in Fig.~\ref{fig:intervention} reflect our physical intuition about the likely influence of an unobserved process on the NAO.
For other applications, different forms might be appropriate, e.g., no tapering, non-symmetric tapering, non-linear tapering, etc.
We recommend keeping $0 \leq \lambda_t \leq 1$, in order to make the coupling effect easily interpretable.
The only other restriction is that the intervention should be transient, not permanent.
Permanent changes can be modelled in the same way, but the effects should be fixed in order to be identifiable, i.e., $\varphi = 1$ and $W_\delta = 0$.

\section{State-space form and prior assessment}
\label{sec:ss}

The model proposed in Sec.~\ref{sec:tvar} can be written in state space form as
\begin{align*}
  Y_t       & = f ( \btheta_t , v_t )       &  
        v_t & \sim \Np{0}{V}                  \\
  \btheta_t & = g ( \btheta_{t-1} , \bw_t ) &
      \bw_t & \sim \Np{\bzero}{\bW}             
  \end{align*}
for $t = 1,\ldots,T$ with $\btheta_0 \sim \Np{\bmu_0}{\bSigma_0}$, where
$\btheta_t = (\mu_t,\beta_t,\allowbreak \psi_{1 t},\psi_{1 t}^\star,\ldots,\psi_{K t},\psi_{K t}^\star,\allowbreak X_{t},\ldots,X_{t-P+1},\allowbreak \phi_{1 t},\ldots,\phi_{P t})^\prime$ and $\bw_t = (w_{\mu t}, w_{\beta t},\allowbreak w_{\psi_1 t}, w_{\psi_1 t}^\star, \ldots, w_{\psi_K t}, w_{\psi_K t}^\star,\allowbreak w_{X t},\allowbreak w_{\phi_1 t}, \ldots, w_{\phi_P t})^\prime$.
The forecast function $f(\btheta_t, v_t)$ is given by (\ref{eqn:observation}).
The evolution function $g(\btheta_{t-1}, \bw_t)$ is given by (\ref{eqn:evolution}).
The evolution covariance matrix $\bW$ is diagonal with main diagonal $\bW_t = (W_\mu, W_\beta,\allowbreak W_\psi, W_\psi,\ldots,W_\psi, W_\psi,\allowbreak W_{X t},\allowbreak W_\phi, \ldots, W_\phi)^\prime$.
The coupling effect $\delta_t$ or effects $\delta_{p t}$ ($p=1,\ldots,P$) can be appended to the state vector $\btheta_t$.
The evolution process vector $\bw_t$ and covariance matrix $\bW$ can also be extended to include the coupling effect evolution process $w_{\delta t}$ or processes $w_{\delta_p t}$ ($p=1,\ldots,P$) and variance $W_\delta$ respectively. 

The prior distribution $\btheta_0 \sim \Np{\bm_0}{\bSigma_0}$ specifies our beliefs about the state-variables at time $t=0$.
We also need to specify priors for the collection of hyper-parameters $\bPhi = (V, W_\mu, W_\beta, W_\psi, W_\phi, W_X, a, b, \alpha, \gamma, \rho, \varphi, W_\delta)^\prime$.

\subsection{Priors for the state variables}
\label{sec:priors}

Independent normal priors were assigned to each component of the state vector $\btheta$ at time $t=0$.
The prior means and variances are listed in Tab.~1.
We were able to use previous studies of the NAO to define informative priors for the mean $\mu_0$ \citep{Hsu1976}, seasonal components $\psi_{1,0},\psi_{1,0}^\star,\psi_{2,0},\psi_{2,0}^\star$ \citep{Chen2012} and TVAR coefficients $\phi_{1,0},\ldots,\phi_{5,0}$ \citep{Masala2015}.
The prior on the local trend $\beta_0$ is based on our judgement that the NAO mean is very unlikely to experience a local change equivalent to more than \SI{1}{\hPa\per\year}.
The daily NAO in Fig.~\ref{fig:nao} has a range of approximately \SI{40}{\hPa}.
Therefore, the TVAR residuals $X_{-4},\ldots,X_0$ were assigned independent normal priors with mean \SI{0}{\hPa} and standard deviation \SI{10}{\hPa}, based on a range of four standard deviations.

In Fig.~\ref{fig:nao}, the NAO index has an inter-annual standard deviation of \SIrange{5}{6}{\hPa} between December and March.
Therefore, in the model with a mean intervention, the coupling effect $\delta_0$ was assigned a normal prior with mean \SI{0}{\hPa} and standard deviation \SI{5}
{\hPa}.
The partial-autocorrelation functions (not shown) for Dec--Mar and Apr--Nov suggest that the change in the autocorrelation structure represented by the coefficients $\phi_{1,0},\ldots,\phi_{P t}$ is quite small.
Therefore, in the model with an intervention on the autocorrelation structure, the coupling effects $\delta_{1,0},\ldots,\delta_{P,0}$ were assigned normal priors with mean \SI{0.0}{\hPa} and standard deviation \SI{0.2}{\hPa}.

\begin{table}
  \caption{Prior probability distributions for the state variables $\btheta_0$.
           All normally distributed.
           \label{tab:priors}}
  \centering
  \begin{small}
    \begin{tabular}{lccc}
      \hline
      Component & Parameter & Mean & Variance \\
      \hline
      Mean level            & $\mu_0$                
                            & \SI{6.0}{\hPa}             & $1^2$      \\
      Local trend           & $\beta_0$              
                            & \SI{0.0}{\hPa\per\year}    & $0.002^2$  \\
      Annual cycle          & $\psi_{1,0}$ 
                            & \SI{3.6}{\hPa}             & $1.0^2$    \\
      Annual cycle          & $\psi_{1,0}^\star$
                            & \SI{1.0}{\hPa}             & $1.5^2$    \\
      Semi-annual cycle     & $\psi_{2,0}$               
                            & \SI{1.3}{\hPa}             & $0.9^2$    \\
      Semi-annual cycle     & $\psi_{2,0}^\star$               
                            & \SI{0.7}{\hPa}             & $1.3^2$    \\
      Irregular component   & $X_{-4},\ldots,X_0$ 
                            & \SI{0.0}{\hPa}             & $10^2$     \\
      TVAR coefficients     & $\phi_{1,0},\ldots,\phi_{5,0}$
                            & $+1.8,-1.3,+0.7,-0.3,+0.1$ & $0.2^2$    \\        
    \end{tabular}
  \end{small}
\end{table}

\subsection{Priors on the hyper-parameters}
\label{sec:variances}

The prior distributions assigned to the hyper-parameters $V$, $W_\mu$, $W_\beta$, $W_\phi$, and $W_X,a,b$ are listed in Tab.~\ref{tab:variances}.
In the case of the NAO, the variability in the mean and seasonal components will be driven primarily be solar forcing, therefore we assume equal error variances, i.e., $W_\psi = W_\mu$.
The observation and evolution variances $V$, $W_\mu$, $W_\beta$ and $W_\phi$ are all expected to be very small, but non-zero.
Therefore, boundary-avoiding priors were specified in the form of Normal distributions on the log of each variance parameter.
Simulation studies of the individual components in (\ref{eqn:evolution}) were used to assign priors that reflect the range of variability we consider plausible for each component.
We expect the annual cycle in the day-to-day variance $W_{X t}$ to peak during the winter season (Dec-Jan-Feb) with an amplitude of up to \SI{5}{\hPa\squared}.
Corresponding uniform priors were assigned to the amplitude and phase of $W_{X t}$, and transformed to approximate normal priors for $a$ and $b$ by simulation. 

\begin{table}
  \caption{Prior densities for hyper-parameters.
           \label{tab:variances}}
  \centering
  \begin{small}
    \begin{tabular}{lccc}
      \hline
      Component & Parameter & Prior & $\approx \SI{95}{\percent}$ Interval \\
      \hline
      Observation variance & $\log V$        & $\Np{-10}{3^2}$ & $(-16,- 4)$ \\
      Mean variance        & $\log W_\mu$    & $\Np{-12}{3^2}$ & $(-18,- 6)$ \\     
      Trend variance       & $\log W_\beta$  & $\Np{-28}{3^2}$ & $(-34,-22)$ \\     
      Irregular variance   & $\log W_X$      & $\Np{0.0}{1^2}$  & $(-2.0,+2.0)$   \\
      Irregular variance   & $a$             & $\Np{0.5}{1^2}$ & $(-1.5,+2.5)$   \\
      Irregular variance   & $b$             & $\Np{2.0}{1^2}$ & $( 0.0,+4.0)$   \\     
      Coefficient variance & $\log W_\phi$   & $\Np{-18}{3^2}$ & $(-24,-12)$ \\
      \hline
    \end{tabular}
  \end{small}
\end{table}

Table~3 lists the priors for the intervention parameters $\alpha,\gamma$ and $\rho$ and the coupling effect parameters $\varphi$ and $W_\delta$.
Our beliefs about the timing of the intervention $\lambda_t$ are the same regardless of whether coupling effects the mean or the autocorrelation structure.
Vague triangular priors are specified for the beginning $\alpha$ and duration $\gamma$ of the coupled period.
These suggest a coupled period with total length around 180~days, beginning around 1 November.
A mildly informative prior is specified for the tapering parameter $\rho$ to reflect our physical reasoning that the influence of the unobserved process is unlikely to be constant throughout the coupled period.
The coupling coefficients $\varphi_\mu$ and $\varphi_X$ are expected to be positive and close to but not exceeding one.
The mean coupling effect variance $W_{\delta_\mu}$ is expected to be greater than the mean variance $W_\mu$, but still small compared to $W_{X_t}$. 
Similarly, the autocorrelation coupling effect variance $W_{\delta_X}$ is expected to be greater than the coefficient evolution variance $W_\phi$.

\begin{table}
  \caption{Prior densities for intervention hyper-parameters.
           \label{tab:intervention}}
  \centering
  \begin{small}
    \begin{tabular}{lccc}
      \hline
      Component & Parameter & Prior & $\approx \SI{95}{\percent}$ Interval \\
      \hline
      Coupling start       & $\alpha - 120$        & $\Tri (0, 365, 185)$   & $(40,325)$ \\
      Coupling length      & $\gamma$        & $\Tri (0,365,180)$   & $(40,325)$ \\
      Tapered proportion   & $\rho$          & $\Bp{4}{6}$     & $(0.15,0.70)$   \\
      Mean effect variance & $\log W_{\delta_\mu}$ & $\Np{- 8}{4^2}$ & $(-16,  0)$ \\
      Mean effect coefficient   & $\varphi_\mu$       & $\Bp{4}{1}$     & $(0.4,1.0)$   \\
      Autocorrelation effect variance & $\log W_{\delta_\phi}$ & $\Np{-16}{4^2}$ & $(-24,- 8)$ \\
      Autocorrelation effect coefficient   & $\varphi_\phi$       & $\Bp{45}{1}$     & $(0.9,1.0)$   \\
      \hline
    \end{tabular}
  \end{small}
\end{table}

\section{Posterior Inference}
\label{sec:inference}

We want to evaluate the joint posterior of the model components $\btheta_1,\ldots,\btheta_T$ and the hyper-parameters $\bPhi$
\begin{align*}
  \Prp{\btheta_{1:T}, \bPhi \mid Y_{1:T}} 
    & = \Prp{\btheta_{1:T} \mid \bPhi,Y_{1:T}}
        \Prp{\bPhi \mid Y_{1:T}}.
\end{align*}

If both $f(\btheta_t, v_t)$ and $g(\btheta_{t-1}, \bw_t)$ were linear functions, then conditional on $\bPhi$, we could sample from the marginal posterior of the state variables $\Prp{\btheta_{1:T} \mid \bPhi,Y_{1:T}}$
using the well known \emph{forward-filtering backward-sampling} algorithm \citep{Fruhwirth1994}.
However, the evolution function $g(\btheta_{t-1}, \bw_t)$ defined by (\ref{eqn:evolution}) is non-linear due to the combination of $\phi_p$ and $X_{t-p}$ in (\ref{eqn:evolution}e).
The form of the observation equation in (\ref{eqn:fdeltap}) is also contains a non-linear combination of $\delta_{p t}$ and $X_{t-p}$.
Therefore, we use linear approximations of the observation and state equations 
\begin{align*}
  Y_t       & \approx f \left( \hat{\btheta}_t, \hat{v}_t \right)
              + \frac{\partial f}{\partial \btheta} 
                \Bigr\rvert_{\hat{\btheta}_t,\hat{v}_t}
                \left( \btheta_t - \hat{\btheta}_t \right)
              + \frac{\partial f}{\partial v} 
                \Bigr\rvert_{\hat{\btheta}_t,\hat{v}_t}
                \left( v_t - \hat{v}_t \right) \\
  \btheta_t & \approx g   \left( \hat{\btheta}_{t-1}, \hat{\bw}_t \right)
              + \frac{\partial g}{\partial \btheta} 
                \Bigr\rvert_{\hat{\btheta}_{t-1},\hat{\bw}_t}
                \left( \btheta_{t-1} - \hat{\btheta}_{t-1} \right)
              + \frac{\partial g}{\partial \bw} 
                \Bigr\rvert_{\hat{\btheta}_{t-1},\hat{\bw}_t}
                \left( \bw_t - \hat{\bw}_t \right)
\end{align*}
where $\hat{\btheta}_{t-1} = \Ep{\btheta_{t-1}}$, $\hat{\btheta}_t = \Ep{\btheta_t}$, $\hat{\bw}_t = \Ep{\bw_t}$ and $\hat{v}_t = \Ep{v_t}$.
This linearisation leads to approximate forward-filtering backward-sampling recursions, detailed in Appendix~\ref{app:ss}.

In general, we expect the TVAR evolution variance $W_{\bphi}$ to be be small, so the coefficients $\phi_{1 t},\ldots,\phi_{P t}$ will be only weakly correlated with the other state variables and our uncertainty about the coefficients will decrease rapidly over time.
Since the other components of the evolution function $g(\btheta_{t-1},  \bw_t)$ are linear and the observation errors $v_t$ and joint state evolution process $\bw_t$ are normal, forward-filtering and backward-sampling based on the linear approximation is expected to be very accurate.
Simulation study showed that the linear approximation provides excellent filtering and smoothing performance, even when all components of the model evolve much more rapidly than expected (see supplementary material).
The linear approximation sometimes struggles to distinguish the TVAR coefficients $\phi_{1 t},\ldots,\phi_{P t}$ from the intervention effects $\delta_{1 t},\ldots,\delta_{P t}$ in the autocorrelation intervention model when  both sets of coefficients evolve rapidly.
In the case of the NAO, we expect only slow evolution of the TVAR coefficients, and little of no change in the intervention effects.
In this scenario, the linearised approximation performs very well.

The marginal posterior of the hyper-parameters $\Prp{\bPhi \mid Y_{1:T}}$ is proportional to
\begin{align*}
  \Prp{\bPhi \mid Y_{1:T}} 
    & \propto \Prp{Y_{1:T} \mid \bPhi} \Prp{\bPhi}.
\end{align*}
The marginal likelihood $\Prp{Y_{1:T} \mid \bPhi}$ can be decomposed as
\begin{align*}
  \Prp{Y_{1:T} \mid \bPhi}
    & = \Prp{Y_1 \mid \bPhi} \prod_{t=2}^T \Prp{Y_t \mid Y_{1:t-1}, \bPhi}.
\end{align*}
The forward-filtering recursions in Appendix~A include an expression for the one-step ahead forecast distribution $\Prp{Y_t \mid Y_{1:t-1}, \bPhi}$.
So the marginal likelihood can be evaluated analytically.
Therefore, the joint posterior $\Prp{\btheta_{1:T}, \bPhi \mid Y_{1:T}}$ can be efficiently sampled by combining forward-filtering backward-sampling with a Metropolis-Hastings scheme targeting $\Prp{\bPhi \mid Y_{1:T}}$ as follows
\begin{itemize}
 \item Let $j$ denote a sample index, at $j = 1$
   \begin{itemize}
     \item Sample starting values $\bPhi^{(1)}$;
     \item Sample $\btheta_{1:T}^{(1)} \mid \bPhi^{(1)},Y_{1:T}$ by backward-sampling.
   \end{itemize}
 \item For $j = 2,\ldots,J$
  \begin{itemize}
    \item Sample new values $\bPhi^\star$ from proposal $q(\bPhi^\star \mid \bPhi)$ ;
    \item Accept $\bPhi^\star$ with probability
          \begin{equation*}
            \min \left\lbrace \frac
              {\Prp{Y_{1:T} \mid \bPhi^\star} \Prp{\bPhi^\star} q(\bPhi \mid \bPhi^\star)}
              {\Prp{Y_{1:T} \mid \bPhi} \Prp{\bPhi} q(\bPhi^\star \mid \bPhi)}
              , 1 \right\rbrace;
          \end{equation*}
    \item Sample $\btheta_{1:T}^{(j)} \mid \bPhi^{(j)},Y_{1:T}$ by backward-sampling.
  \end{itemize}
\end{itemize}
In practice, it is not necessary to perform backward-sampling for the state $\btheta_{1:T}$ for every sample $(j)$.
As with any Markov Chain Monte Carlo approach, there is likely to be significant autocorrelation between subsequent samples of the hyper-parameters $\bPhi^{(j)}$.
In the interest of saving storage and computation time, it is sufficient to sample the state $\btheta_{1:T}$ for a subset of the $\bPhi^{(j)}$.

\subsection*{Alternative approaches}

Conditional on the TVAR coefficients $\phi_{1 t},\ldots,\phi_{P t}$, the model defined by (\ref{eqn:observation}) and (\ref{eqn:evolution}) is a normal dynamic linear model.
We could split the state vector $\btheta_t$ into two parts $\btheta_t^\star = (\mu_t,\beta_t,\allowbreak \psi_{1 t},\psi_{1 t}^\star,\ldots,\psi_{K t},\psi_{K t}^\star,\allowbreak X_{t},\ldots,X_{t-P+1})^\prime$ and $\bphi_t^\star = (\phi_{1 t},\ldots,\phi_{P t})^\prime$, and then alternate between forward-filtering and backward-sampling for each part, conditional on the other.
Gibbs' sampling steps could be used to sample the hyper-parameters $\bPhi$ \citep[Chapter 15]{WestHarrison}.
This approach provides exact sampling from the required posterior distribution, but has two drawbacks compared to the approximate approach proposed here.
First, backward sampling must be performed at every iteration, making this approach computationally expensive.
Second, Gibbs' sampling based on the full conditional distributions of the hyper-parameters will tend to mix very slowly, especially for long time series where the data completely overwhelm the prior.

Particle filtering methods provide tools for inference in general non-linear and non-normal state-space models \citep{Doucet2011}.
However, particle filters are computationally expensive and suffer from problems of ``particle degeneracy'', i.e., the state $\btheta_t$ will eventually be represented by a single particle at times $t \ll T$.
Since we are interested in what happened at all times $t = 1,\ldots,T$, we also require particle smoothing in order to overcome the degeneracy problem \citep{Godsill2004,Briers2010}.
Particle smoothing is even more computationally expensive, making an alternative approach highly desirable.
The problem of efficient inference for unknown hyper-parameters also remains an active topic for research in Sequential Monte Carlo methods \citep{Chopin2013}.

\subsection{Model selection}

For some applications, it will be possible to choose between the mean and autocorrelation intervention models on the basis of posterior predictive diagnostics, i.e., whether the model reproduce the observed behaviour.
The posterior distributions of the hyper-parameters $\bPhi$ can also be useful for choosing between models, e.g., is the coupling effect variance $W_\delta$ negligible.
More formally, we can compare the two intervention models by evaluating the Bayes' factor
\begin{align}
  B = \frac{\Prp{Y_{1:T} \mid M_\mu}}{\Prp{Y_{1:T} \mid M_X}}
    = \frac{\int \Prp{Y_{1:T} \mid \bPhi, M_\mu} \Prp{\bPhi \mid M_\mu} d \bPhi}
           {\int \Prp{Y_{1:T} \mid \bPhi, M_X  } \Prp{\bPhi \mid M_X  } d \bPhi}
  \label{eqn:bf}
\end{align}
where $M_\mu$ is the model including an intervention on the mean, and $M_X$ is the model including and intervention on the temporal dependence structure.
The Bayes' factor is defined as the ratio of the marginal likelihoods of the competing models \citep{Kass1995}.
Values of $B$ greater than one indicate support for the mean intervention model $M_\mu$ and values of $B$ less than one indicate support for the autocorrelation intervention model $M_X$.
The conditional likelihoods $\Prp{Y_{1:T} \mid \bPhi, M}$ can be evaluated using the filtering recursions in Appendix~A.
The marginal likelihoods $\Prp{Y_{1:T} \mid M}$ can be evaluated based on the posterior samples $\bPhi^{(j)} \mid Y_{1:T}, M$ ($j=1,\ldots,J$) by bridge sampling \citep{Gronau2017}.

\subsection{What is the effect of the coupling?}

Given posterior samples $\btheta_{1:T}^{(j)} \mid \bPhi^{(j)},Y_{1:T},M$, we can make inferences about any function of the state variables $\btheta_t$ for any time period $\boldsymbol{\tau}$ of interest, e.g., $\boldsymbol{\tau} = \lbrace \text{Dec 2009--Mar 2010} \rbrace$.
It is useful to define $\eta_t = \mu_t + \sum_k \psi_{k t}$, which we refer to as the systematic component of the observed process.
The relative contributions to the variability between coupled periods of the systematic component $\eta_t$, the irregular component $X_t$, the coupling effects $\delta_t$ or $\delta_{p t}$ ($p=1,\ldots,P$) and observation error $v_t$ are of particular interest.
The means of the systematic and irregular components during period $\boldsymbol{\tau}$ in the $j$th sample are
\begin{align}
  \bar{\eta}_{\boldsymbol{\tau}}^{(j)} & = \frac{1}{n} \sum_{t \in {\boldsymbol{\tau}}} \eta_t^{(j)} & \text{and} & &
  \bar{X}_{\boldsymbol{\tau}}^{(j)} & = \frac{1}{n} \sum_{t \in {\boldsymbol{\tau}}} X_t^{(j)}
  \label{eqn:tbars}
\end{align}
where $n$ is the number of time steps in $\boldsymbol{\tau}$.
The means of the coupling effects during period $\boldsymbol{\tau}$ in the mean and autocorrelation models respectively are
\begin{align}
  \bar{\delta}_{\mu \boldsymbol{\tau}}^{(j)} & = \frac{1}{n} \sum_{t \in {\boldsymbol{\tau}}} \lambda_t \delta_t^{(j)} & \text{and} & &
  \bar{\delta}_{X \boldsymbol{\tau}}^{(j)} & = 
    \frac{1}{n} \sum_{t \in {\boldsymbol{\tau}}} \lambda_t 
      \sum_{p=1}^P \delta_{p t}^{(j)} X_{t-p}^{(j)}.
  \label{eqn:dbars}
\end{align}
The contribution due to observation error is
\begin{align*}
  \bar{v}_{\boldsymbol{\tau}}^{(j)} 
    & = \bar{Y}_{\boldsymbol{\tau}}
      - \bar{\eta}_{\boldsymbol{\tau}}^{(j)}
      - \bar{\delta}_{\boldsymbol{\tau}}^{(j)}
      - \bar{X}_{\boldsymbol{\tau}}^{(j)}
\end{align*}
where $\bar{Y}_{\boldsymbol{\tau}} = \sum_{t \in \boldsymbol{\tau}} Y_t / n$.
The prior expectations of the irregular component $X_t$ and the coupling effects $\delta_t$ or $\delta_{p t}$ ($p=1,\ldots,P$) during any period $\boldsymbol{\tau}$ are zero by (\ref{eqn:evolution}e), (\ref{eqn:delta}) and (\ref{eqn:deltap}), i.e., $\Ep{X_t} = 0$ and $\Ep{\delta_t} = \Ep{\delta_{p t}} = 0$ for all $t$.
In general $\Ep{\eta_t} \neq 0$, so for the systematic component $\eta_t$ it is more useful to consider the anomalies over all similar periods
\begin{align*}
  \bar{\eta}_{\boldsymbol{\tau}}^{\star (j)} = 
    \bar{\eta}_{\boldsymbol{\tau}}^{(j)}  - 
      \frac{1}{\vert D \vert} \sum_{t^\prime \in D} \eta_{t^\prime}^{(j)}
\end{align*}
where $D = \lbrace t \in 1,\ldots,T : d(t) = d(s) \ \text{and} \ s \in {\boldsymbol{\tau}} \rbrace$ and $d(t)$ is the day of the year at time $t$.
The sample means
\begin{align}
  \bar{\eta}_{\boldsymbol{\tau}}^\star 
    & = \frac{1}{J} \sum_j \bar{\eta}_{\boldsymbol{\tau}}^{\star (j)}, & 
  \bar{\delta}_{\boldsymbol{\tau}} 
    & = \frac{1}{J} \sum_j \bar{\delta}_{\boldsymbol{\tau}}^{(j)}, & 
  \bar{X}_{\boldsymbol{\tau}} 
    & = \frac{1}{J} \sum_j \bar{X}_{\boldsymbol{\tau}}^{(j)}, & 
  \bar{v}_{\boldsymbol{\tau}} 
    & = \frac{1}{J} \sum_j \bar{v}_{\boldsymbol{\tau}}^{(j)}
  \label{eqn:means}
\end{align}
provide a summary of the posterior expected contribution of each component during the period $\boldsymbol{\tau}$.
Quantiles can also be computed over the samples to form credible intervals for the contribution of each component.

\subsection{Analysis of variance}

In a stationary model, elements of the marginal posterior $\Prp{\bPhi \mid Y_{1:T},M}$ would summarise the relative contributions of each model component to the observed variability in the index $Y_t$.
However, since our model is non-stationary, we require an alternative summary of the variance components.
In particular, we are interested in the proportion of the inter-annual variance of the winter (Dec-Jan-Feb) mean of the NAO index explained by each component.
Let $\boldsymbol{\tau}_i$ be the i$th$ winter period.
We propose performing an analysis of variance of the observed means $\bar{y}_{\tau_i} = \frac{1}{N_i} \sum_{t \in \tau_i} y_t$ for each sample $j$ using the component means $\bar{\eta}_{\boldsymbol{\tau}_i}^{(j)}$, $\bar{\delta}_{\boldsymbol{\tau}_i}^{(j)}$ and $\bar{X}_{\boldsymbol{\tau}_i}^{(j)}$ defined in (\ref{eqn:tbars}) and (\ref{eqn:dbars}) as explanatory variables.
The analysis of variance leads to four sums-of-squares for each sample $j$, corresponding to the sum of squared deviations explained by the systematic $\eta_t$ and irregular $X_t$ components, the coupling effects $\delta_t$ or $\delta_{p t}$ ($p=1,\ldots,P$) and observation errors $v_t$ in each sample trajectory.
Posterior summaries over the $J$ samples summarise the overall contributions of each component to the variability between coupled periods.

\subsection{Can we make predictions using unobserved components?}

Knowledge of the unobserved component through the intervention effect $\delta_t$ should provide useful predictability within coupled periods.
The model proposed in Sec.~\ref{sec:coupling} also allows for dependence between successive coupled periods, so knowledge of the unobserved component during one coupled period may also be useful for predicting the next.
The $k$-step ahead forecast distribution given data up to time $t$ can be sampled exactly using the recursions in Appendix~\ref{app:ss}.
The correlation between the data and the forecast means provides a simple measure of forecast performance.

\section{Results}
\label{sec:results}

The Metropolis-Hastings sampler outlined in Sec.~\ref{sec:inference} was used to draw \num{1000} samples from each of the joint posteriors $\Prp{\btheta_{1:T}, \bPhi \mid Y_{1:T}, M_\mu}$ and $\Prp{\btheta_{1:T}, \bPhi \mid Y_{1:T}, M_X}$.
Full details of the sampling design, proposal distributions, diagnostic trace plots and posterior density plots are given in the supplementary material.
Both models converge to stable distributions and mix efficiently, however the burn-in period can be very long depending on the initial values of the hyper-parameters $\bPhi$.

\begin{figure}[t]
  \centering
  \includegraphics[width=0.49\textwidth]{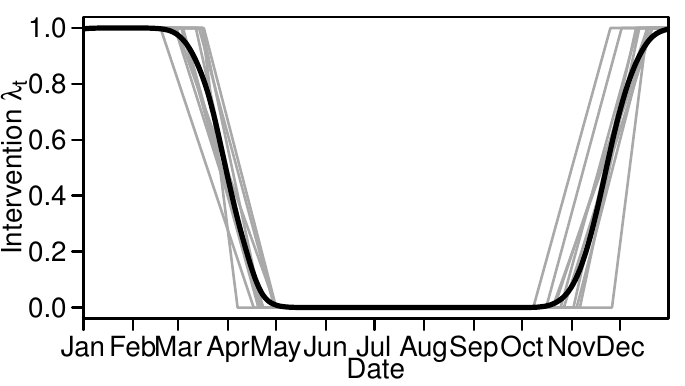}
  \includegraphics[width=0.49\textwidth]{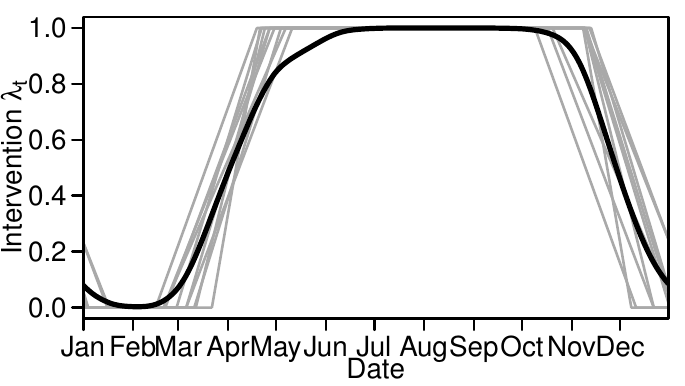}
  \caption{Posterior of the intervention $\lambda_t$.
           (left) The model with an intervention on the mean $M_\mu$;
           (right) the model with an intervention on the autocorrelation structure $M_X$.
           Grey lines represent a random sample of 10 realisations of the intervention $\lambda_t^{(j)}$ based on the posterior samples of $\alpha$, $\gamma$ and $\rho$.
           The black line is the pointwise posterior mean over all \num{1000} realisations of $\lambda_t^{(j)}$.
           \label{fig:vispost}}
\end{figure}

Despite deliberately vague prior distributions, the posterior distributions of the intervention parameters $\alpha$ and $\gamma$ are quite sharp for both models.
Figure~\ref{fig:vispost} visualises the posterior distribution of the intervention $\lambda_t$ for each model.
In the mean intervention model $M_\mu$, an unobserved component acts strongly on the NAO between December and February and into March.
There is almost no evidence of coupling between May and October.
In the autocorrelation intervention model $M_X$ the situation is reversed.
The inverted intervention structure is unexpected, but still consistent with a marked difference in behaviour between the extended winter (Dec--Mar) and extended summer (Apr--Nov) seasons.
Prior sensitivity analysis showed that the posterior distributions of the hyper-parameters $\bPhi$, including $\alpha$ and $\gamma$, are insensitive to the choice of priors in Tables~\ref{tab:variances} and \ref{tab:intervention} (see supplementary material).

In order to assess the identifiability of the various model components, particularly the coupling effects, we computed correlation matrices for the states $\btheta_{1:T}^{(j)} \mid \bPhi^{(j)}, \bY_{1:T}$ for each sample $j$.
On average across the \num{1000} sample covariance matrices, the state variables in both models are all uncorrelated with one another.
In particular, the mean intervention effect $\delta_t$ is almost completely uncorrelated with the irregular component $X_t$ (\SI{90}{\percent} CI of correlation $(-0.02,+0.06)$), and only ever weakly correlated with the mean component $\mu_t$ (\SI{90}{\percent} CI $(-0.28,+0.20)$).
While the autocorrelation intervention effects $\delta_{1 t},\ldots,\delta_{5 t}$ are uncorrelated with the other state variables on average, they may be strongly correlated \emph{or} anti-correlated with the mean $\mu_t$ and the autocorrelation coefficients $\phi_{1 t},\ldots,\phi_{5 t}$.
Further investigation showed that these strong associations were the result of the slow rate of change of these parameters, since sampling multiple state trajectories $\btheta_{1:T}$ from any single sample of the hyper-parameters $\bPhi_{(j)}$ produced a similar range of sample correlations.

\begin{figure}[t]
  \includegraphics[width=0.49\textwidth]{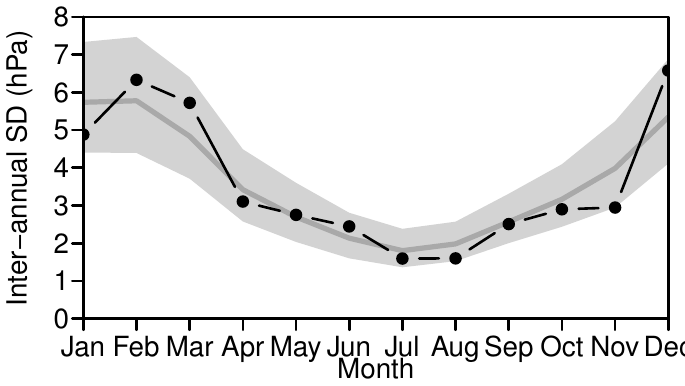}
  \includegraphics[width=0.49\textwidth]{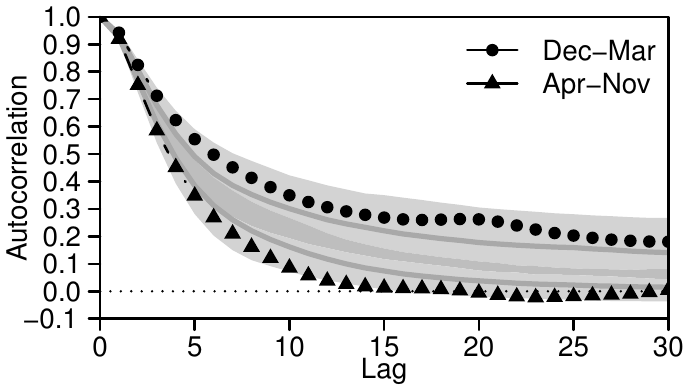} \\
  \includegraphics[width=0.49\textwidth]{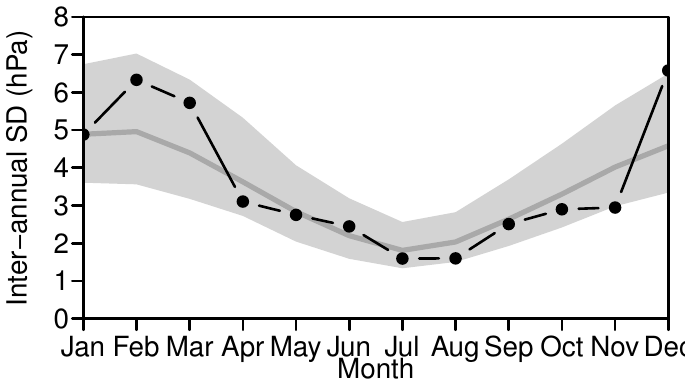}
  \includegraphics[width=0.49\textwidth]{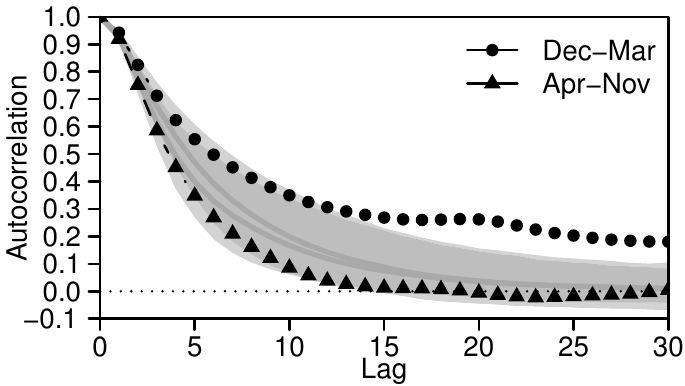}
  \caption{Posterior predictive checks.
           (top) The model with an intervention on the mean $M_\mu$;
           (bottom) the model with an intervention on the autocorrelation structure $M_X$.
           Before computing the autocorrelation, the mean, a linear trend, annual and semi-annual cycles were estimated by least-squares and removed.
           Black lines represent the observed statistics.
           Dark grey lines indicate the posterior mean.
           Shading indicates pointwise \SI{90}{\percent} posterior credible intervals.
           Dark grey shading in (bottom right) indicates overlap between credible intervals.
           \label{fig:checks}}
\end{figure}

Posterior predictive diagnostics were used to check the performance of each model in capturing the observed structure of the NAO.
In particular, we are interested whether the model can reproduce the seasonal contrast in the  inter-annual variance and autocorrelation structures in Fig.~\ref{fig:nao}.
For each sample $\btheta_{1:T}^{(j)}, \bPhi^{(j)} \mid Y_{1:T}$ from each model we simulate a new sequence of states $\btheta_{\boldsymbol{\tau}}^{\star (j)} \mid \bPhi^{(j)}, Y_{1:T}$ and observations $Y_{\boldsymbol{\tau}}^{\star (j)}$ for the period $\boldsymbol{\tau} = \lbrace \text{Jan 1988--Dec 2017} \rbrace$.
Figure~\ref{fig:checks} compares the annual cycle in the inter-annual standard-deviation and the seasonal autocorrelation functions of the observed data $Y_{\boldsymbol{\tau}}$ and the samples $Y^{\star (j)}_{\boldsymbol{\tau}}$ for $j = 1,\ldots,1000$.
The mean intervention model $M_\mu$ is able to reproduce both the inter-annual variability and the seasonal autocorrelation function.
There is a clear difference in the autocorrelation functions simulated between April and November, and between December and March.
However, the autocorrelation intervention model $M_X$ is unable to reproduce the seasonal autocorrelation behaviour and doesn't reproduce the inter-annual variability as well as the mean intervention model $M_\mu$.
There is a small difference between the extended summer (Apr--Nov) and extended winter (Dec--Mar) autocorrelation functions, but much less than observed in the data.
The inverted intervention structure in Fig.~\ref{fig:vispost} is an attempt exploit the extended summer (Apr--Nov) period to distinguish the small intervention effects $\delta_p$.
Similar checks (not shown) suggest that both models are able to adequately capture the annual cycle in the NAO, indicating that our choice of $K=2$ harmonics was reasonable.

The posterior predictive checks strongly favour the mean intervention model over the autocorrelation intervention model.
The mean intervention is able to reproduce the observed behaviour, the autocorrelation intervention cannot.
The Bayes' factor of $B = \num{1096}$ also provides extremely strong evidence in favour of the mean intervention model, i.e., the observed data are almost \num{1000} times more likely to have arisen from the mean intervention model.
We conclude that the most likely explanation for the observed behaviour of the NAO index is a transient change in the mean level during the extended winter (Dec--Mar) season.
The remainder of our analysis focuses on interpreting only the mean intervention model.

\begin{figure}[t]
  \includegraphics[width=0.49\textwidth]{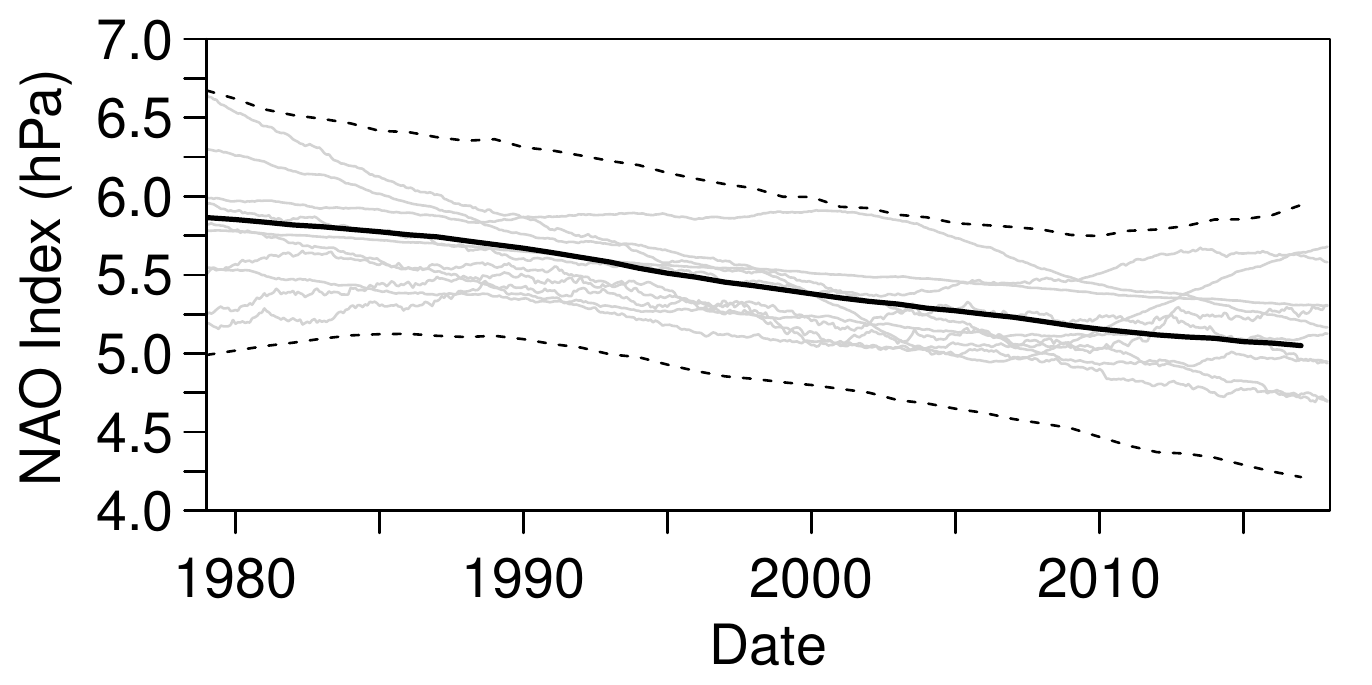}
  \includegraphics[width=0.49\textwidth]{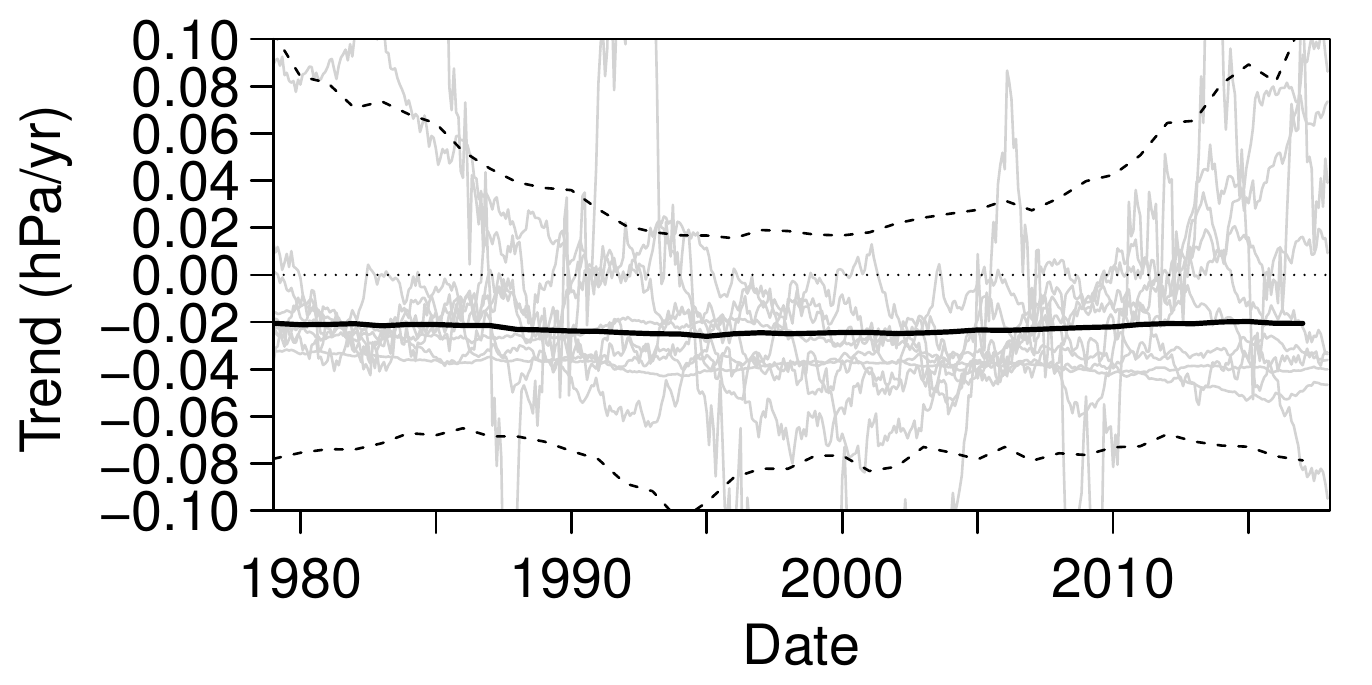} \\
  \includegraphics[width=0.49\textwidth]{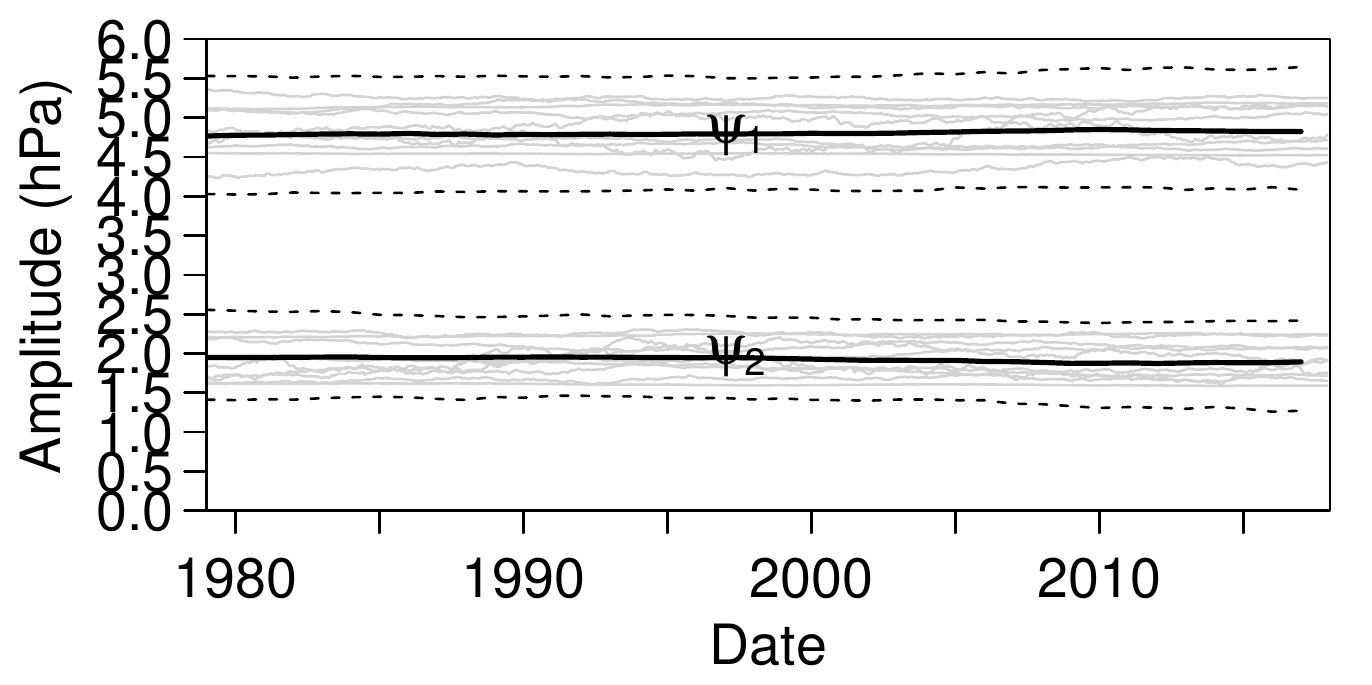}
  \includegraphics[width=0.49\textwidth]{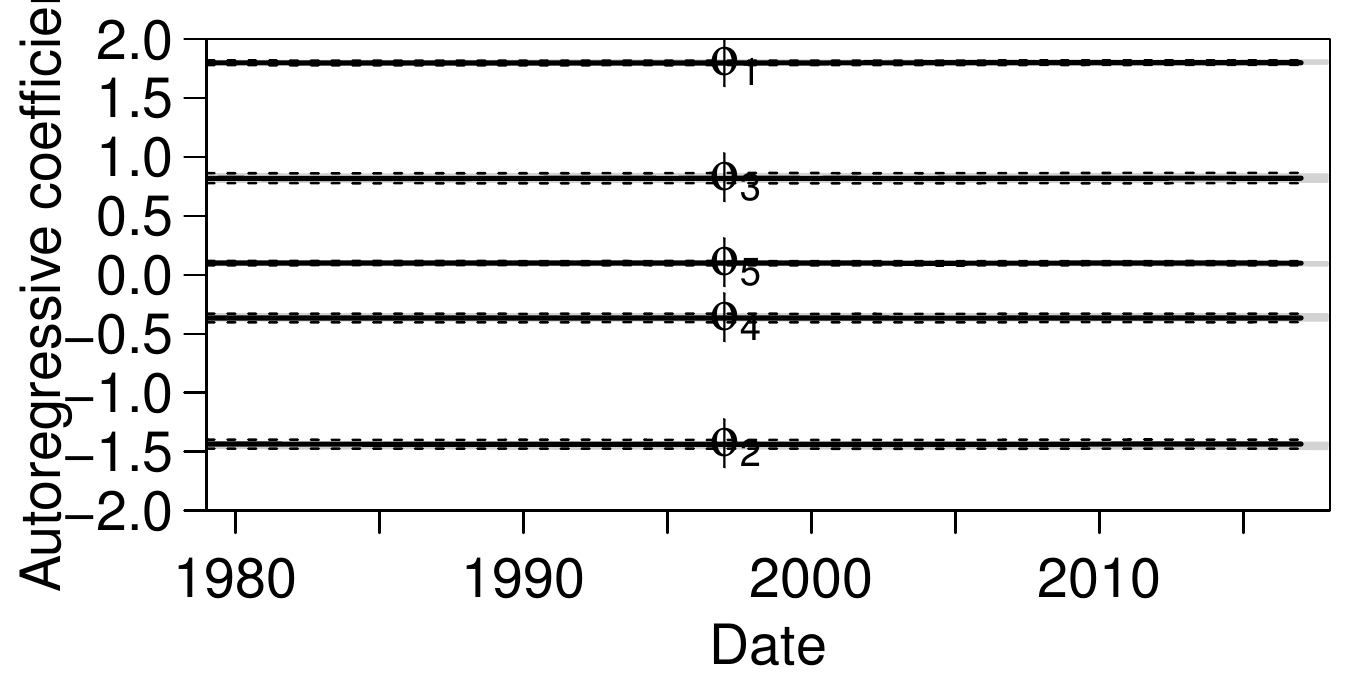}
  \caption{Posterior distributions of model components.
           (top left) Mean $\mu_t$;
           (top right) trend $\beta_t$;
           (bottom left) amplitudes of seasonal harmonics $\psi_1$ and $\psi_2$;
           (bottom right) TVAR coefficients $\phi_{1 t},\ldots,\phi_{5 t}$.
           Solid black lines represent the pointwise posterior mean.
           Dashed black lines represent pointwise \SI{90}{\percent} credible intervals.
           Grey lines are a random sample of 10 trajectories $\btheta_{1:T}^{(j)} \mid \bPhi^{(j)}, Y_{1:T}$.
           \label{fig:trends}}
\end{figure}

Surprisingly for such a complex phenomenon, the mean, trend and seasonal components of the NAO index show very limited evidence of non-stationarity.
Figure~\ref{fig:trends} shows a number of posterior trajectories $\btheta_{1:T}^{(j)}$ from each component.
There is evidence of a fairly constant trend leading to a reduction in the mean level of the NAO of around \SI{0.8}{\hPa} between 1979 and 2017.
The posterior distribution of the trend itself suggests that the rate of decrease in the NAO mean peaked around 1993--94 at around \SI{0.03}{\hPa\per\year} $(-0.07,+0.01)$, since when the trend has gradually weakened.
The amplitudes of the annual and semi-annual cycles are almost constant (likewise the phases).
The 0.95 quantile of the posterior distribution of the mean evolution standard deviation $\sqrt W_\mu$ is \SI{0.005}{\hPa}, so changes in excess of \SI{0.2}{\hPa\per\year} to the mean and seasonal components are not ruled out under the random walk hypothesis.
There is no evidence of non-stationarity in the autoregressive coefficients $\phi_1,\ldots,\phi_5$ which are effectively constant throughout the study period.
This suggests that the day-to-day variation in the NAO can be adequately represented by an AR process rather than a TVAR process.
However, this is a useful conclusion given the observed seasonal autocorrelation structure in Fig.~\ref{fig:nao}.

\subsection*{Quantifying the effect of coupling}

The posterior mean estimate of the intervention effect standard deviation $\sqrt W_{\delta_\mu}$ is 0.43 (0.33--0.53), indicating a very active process, contributing substantial additional inter-annual variability during the extended winter season (Dec--Mar).
Table~4 contains the results of the analysis of variance proposed in Sec.~\ref{sec:inference} for the mean intervention model $M_\mu$.
The effect of coupling $\delta_t$ explains around \SI{66}{\percent} of the observed variation in the winter (Dec-Jan-Feb) means.
Accumulated short-term variability captured by the TVAR residuals $X_t$ explains around \SI{33}{\percent} of the inter-annual variability.
Despite the trend visible in Fig.~\ref{fig:trends}, the contribution of the mean and seasonal components is negligible.
Together they account for a maximum of \SI{5}{\percent} of the inter-annual variability in winter (Dec-Jan-Feb).
In contrast, the mean and seasonal components account for around \SI{15}{\percent} of inter-annual variability in summer (Jun-Jul-Aug) when coupling has no effect and the day-to-day variability is reduced.
The contribution of measurement error is negligible.

\begin{table}
  \caption{Analysis of variance. Bracketed values indicate \SI{90}{\percent} credible intervals.
           \label{tab:anova}}
  \centering
  \begin{footnotesize}
    \begin{tabular}{lcccc}
      \hline
       & Mean & Coupling & Irregular & Error \\
      \hline
       Winter (Dec-Jan-Feb) & 0.00 (0.00,0.05) & 0.66 (0.52,0.77) 
                            & 0.33 (0.23,0.47) & 0.00 (0.00,0.00) \\
       Summer (Jun-Jul-Aug) & 0.15 (0.06,0.22) & 0.00 (0.00,0.00)
                            & 0.85 (0.78,0.94) & 0.00 (0.00,0.00) \\
      \hline
    \end{tabular}
  \end{footnotesize}
\end{table}

\begin{figure}[t]
  \includegraphics[width=\textwidth]{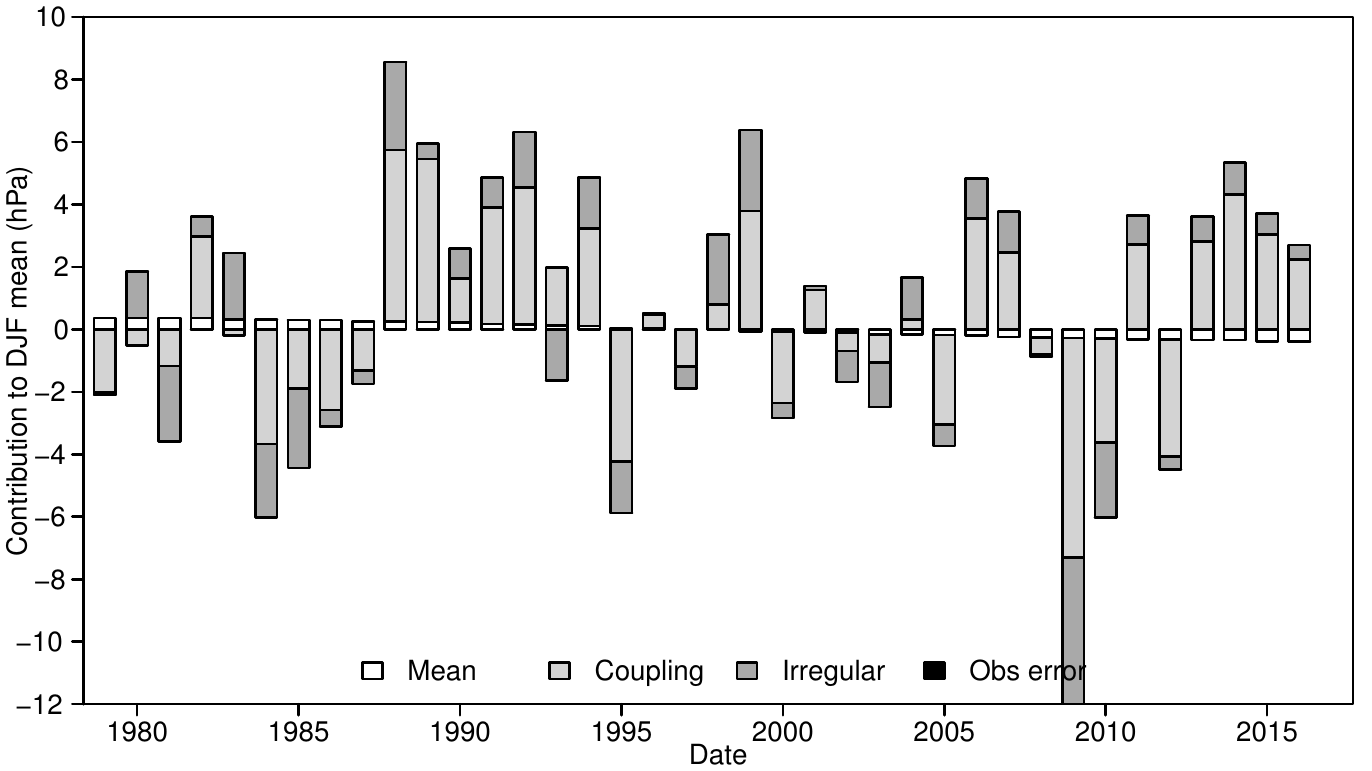}
  \caption{Contribution of individual model components.
           Posterior mean estimates of the winter (Dec-Jan-Feb) mean levels of 
           the systematic component $\bar{\eta}$ ,
           the irregular component $\bar{X}$ ,
           the coupling effect $\bar{\delta}$, and
           the observation error $\bar{v}$.
           \label{fig:attrib}}
\end{figure}

Fig.~\ref{fig:attrib} shows the posterior mean contribution of each component to each observed winter (Dec-Jan-Feb) mean level.
This is an important and useful advance over existing methods in climate science that only estimate the fraction of total variance explained by each component.
The weak negative trend in the mean component $\mu_t$ is clearly visible.
Both the absolute and relative contributions of the irregular component $X_t$ and the coupling effect $\delta_t$ vary between years, but both components usually have the same sign.
This is a product of the limited data available to estimate the components during each extended winter (Dec--Mar).
If the coupling signal cannot be clearly identified during a particular season, then the contribution to the seasonal mean will be split approximately according to the analysis of variance in Tab.~\ref{tab:anova} and the two components will have the same sign.
The fact that the relative contribution of each component varies widely in Fig.~\ref{fig:attrib} indicates that the model is able to separate the coupling effect from the noise of the irregular component.

\subsection*{Forecasting the winter NAO}

\begin{figure}[t]
  \includegraphics[width=0.49\textwidth]{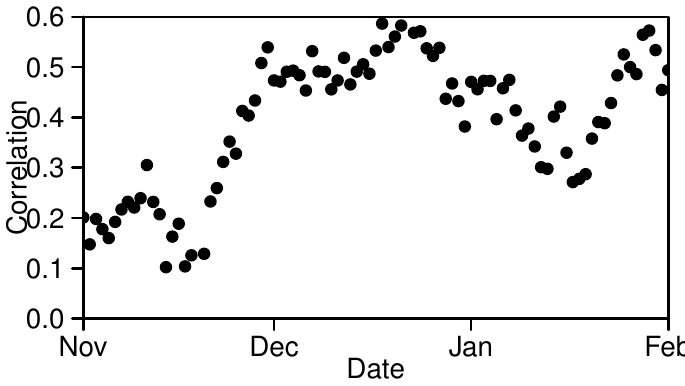}
  \includegraphics[width=0.49\textwidth]{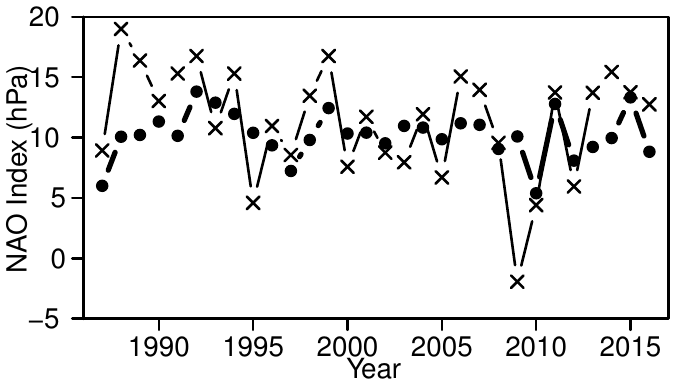}
  \caption{Predictability of winter (Dec-Jan-Feb) NAO.
           (left) The correlation between the observations and forecasts initialised on each day between November and February, for the mean level over the remainder of the period to 28 Feb;
           (right) Observations (\textsf{x}) and forecasts ($\bullet$), for the mean NAO between 1 Dec and 28 Feb each year, given data up to 30 Nov.
           \label{fig:forecasts}}
\end{figure}

The posterior mean estimate of the coupling effect coefficient $\varphi_\mu$ is 0.994 (\SI{90}{\percent} credible interval 0.991--0.997).
In terms of inter-annual variability, this is equivalent to a correlation of around 0.19 (0.05--0.38) between Dec-Jan-Feb means, suggesting limited evidence of persistence and therefore predictability between seasons.
However, if we can learn about the coupling effect quickly enough during a specific coupled period, then we can use that knowledge to provide more skilful forecasts for the rest of the period.
Figure~\ref{fig:vispost} suggests that the system is at least partially coupled from the beginning of November until around the middle of April.
Using the forecasting recursions in Appendix~\ref{app:ss}, we obtained forecasts beginning each day from 1 November to 1 February until the end of the fully coupled period on 28 February for every winter between 1987 and 2016.
Figure~\ref{fig:forecasts} shows the correlation between the forecast and observed means.
By the beginning of December, the correlation approaches 0.5 for the 92 day forecast of the mean NAO to 28 February.
This correlation approaches that achieved by computationally expensive numerical weather prediction models \citep{Scaife2014,Siegert2016}.
The correlation increases slightly as more observations are assimilated during December.
However, as more observations are assimilated, the forecast period decreases and we are essentially predicting weather noise, so the correlation does not increase further.

\begin{figure}[t]
  \centering
  \includegraphics[width=0.49\textwidth]{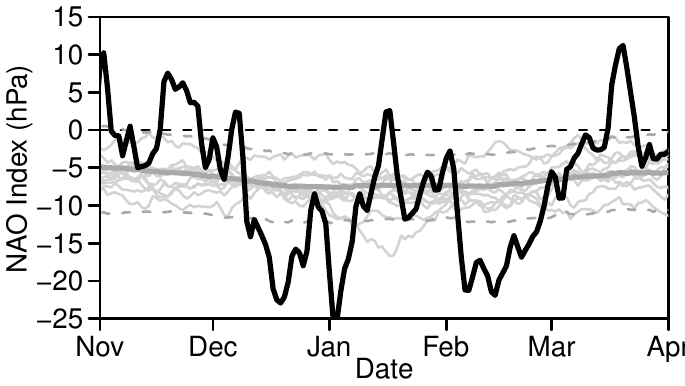}
  \includegraphics[width=0.49\textwidth]{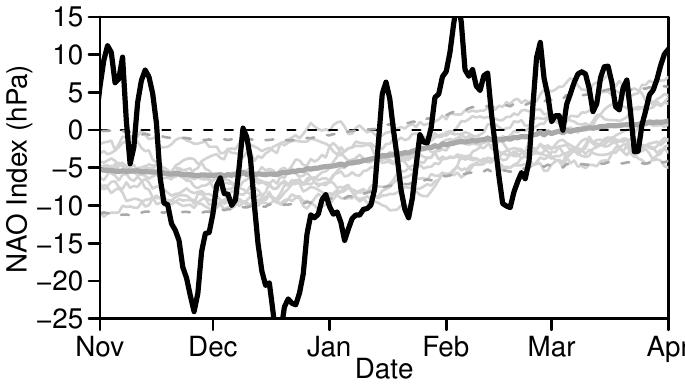}
  \caption{Deseasonalised observations $Y_t - \Ep{\eta_t \mid Y_{1:T}}$ for the winters (Dec-Jan-Feb) of (left) 2009--10 and (right) 2010--11.
           Thin grey lines are a random sample of 10 posterior trajectories for the coupling effect $\delta_t^{(j)} \mid \bPhi^{(j)}, Y_{1:T}$.
           Thick grey and dashed grey lines represent the posterior mean and pointwise \SI{90}{\percent} credible interval for the coupling effect.
           \label{fig:2009-10}}
\end{figure}

Figure~\ref{fig:forecasts} also compares forecasts of the 92 day Dec-Jan-Feb winter mean, initialised on 1 December each year, with the observed mean NAO index for the same periods.
The model predicts the 2010, 2011 and 2012 winter seasons with remarkable accuracy, and captures the general pattern during the 1990s.
However, it fails to predict the extreme winter of 2009--10.
Figure~\ref{fig:2009-10} plots deseasonalised observations of winter 2009--10 ($Y_t - \Ep{\eta_t \mid Y_{1:T}}$).
Deseasonalising the observations leaves only the contributions from the irregular component $X_t$ and the coupling effect $\delta_t$, which represent processes on different time scales.
The irregular component $X_t$ captures high frequency fluctuations, while the coupling effect $\delta_t$ captures any overall departure from the seasonal mean. 
From the middle of December onwards, the mean of the deseasonalised data is clearly negative, which the model attributes to the coupling effect $\delta_t$.
Since the seasonal forecasts in Fig.~\ref{fig:forecasts} were based on information up to 30 November, it is unsurprising that a fairly normal winter was forecast.
In contrast, in winter 2010--11 (Fig.~\ref{fig:2009-10}), a strong negative signal is visible in November which the model is able to exploit to skilfully forecast the remainder of the  Dec-Jan-Feb season.
Winter 2010 also illustrates the time-varying nature of the coupling effect $\delta_t$, which starts strongly negative early in the season, but weakens from mid-January onwards.

\section{Discussion}
\label{sec:discussion}

In this study we have developed Bayesian state space methods for diagnosing predictability in intermittently coupled systems.
Coupling is represented by a transient intervention whose timing and duration is inferred from the data.
Interventions to either the mean or temporal dependence structure are considered.
The effect of intermittent coupling is modelled as dynamic process rather than a sequence of constant and independent effects.
Latent TVAR components are proposed to capture any inherent non-stationarity in the temporal dependence structure.
A linearised approximation is proposed that allows efficient forward-filtering and backward-sampling for models containing latent TVAR components, without requiring complicated and computationally expensive sequential Monte Carlo methods.
In addition, we develop tools for posterior inference in intermittently coupled systems, including evaluating the evidence of a coupling effect, attribution of historical variation in the system, and demonstrating potential predictability.

We applied the proposed model and inference methods to diagnose excess winter-time variability in the North Atlantic Oscillation.
Existing methods in climate science are unable to distinguish between transient changes in the mean or temporal dependence structure.
The proposed model strongly points to transient changes in the mean level of the NAO during a period beginning sometime in November and ending around the middle of April.
This is an important conclusion given that the excess winter-time variability in the NAO is usually characterised by increased temporal dependence.
The mean level of the NAO also appears to change on decadal time scales, in addition to a fairly stable annual cycle and the transient changes in winter-time.
The proposed model is also able to separate the coupling effect from accumulated day-to-day variability in individual seasons.
For the NAO, the two effects actually oppose each other in some seasons.

Like latent AR components, latent TVAR components improve the interpretability of structural time series models by avoiding the need to redefine the mean level of the observed process.
In addition, latent TVAR components permit efficient recursive estimation of the autoregressive parameters and include standard latent AR components as a special case when the evolution variance is zero.
For the NAO, we found little evidence of changes in the autoregressive  structure throughout the study period, so a standard latent AR component could be used to represent day-to-day variability.
However, the fact that we can confirm that autoregressive structure is constant on decadal time-scales is also a useful conclusion.

The model proposed for intermittently coupled systems differs from standard intervention analysis by modelling the effect of repeated coupling events as a dynamic process, rather than a series of independent events.
This allows knowledge gained during one coupled period to inform inferences for the next.
By modelling the coupling effect as a dynamic process, the effect is also able to vary within individual coupled periods rather than being assumed constant.
Climate scientists usually assume that any coupling effect is constant throughout an arbitrarily defined season.
We have shown that the coupling effect on the NAO can vary substantially throughout a single season.

Modelling the effect of coupling as a dynamic process also makes the model robust to minor variations in the timing and duration of the coupled period.
However, the specification of a fixed coupling period remains a limitation.
Hidden Markov and semi-Markov models are widely used in similar seasonal state-switching scenarios to allow for changes in onset and duration \citep[e.g.,][]{Carey-Smith2014}.
Standard hidden Markov models assume instantaneous switching between states.
While such an assumption may be acceptable for some applications, we do not consider it plausible for the NAO.
A completely general alternative would be a reversible-jump MCMC scheme \citep{Green1995}.
In such a scheme, coupling events could be estimated with varying onset, duration or other parametrized structural changes.
However, unless the timing of coupling events varies dramatically, the additional cost and complexity of a reversible-jump scheme seems unnecessary.
The on-line Bayesian changepoint methods proposed by \citet{Fearnhead2011} might provide a more efficient approach.

In the methodology developed here, we have allowed for non-stationarity in the mean and the temporal dependence structure, but not in the variance.
Stochastic volatility models and related ARCH and GARCH models have been widely studied and applied, particularly in economics.
\citet{Masala2015} applied a GARCH model to stochastic modelling of the NAO, but found that its performance was poor.
Efficient filtering and smoothing is possible for time-varying observation error variance \citep[Chapter 10.8]{WestHarrison}.
However, fully conjugate models that admit analytic filtering and smoothing for time-varying state evolution variances are not possible, even in the linear normal case.
Of particular interest are changes in the residual TVAR evolution variance $W_{X t}$ that drives short-term variability in the system.
Sequential Monte Carlo methods or further approximations are required to model time-varying evolution variances.

\vspace{\baselineskip}
\thanks{The authors gratefully acknowledge the support of the Natural Environment Research Council grant NE/M006123/1.
We also wish to thank two anonymous reviewers and an associate editor for their helpful comments.}

\appendix

\section{Forward-filtering, backward-sampling and forecasting}
\label{app:ss}

\subsection*{Forward-filtering}

The sequence of posterior distributions $\lbrace \btheta_t \mid Y_{1:t}, \bPhi : t = 1,\ldots,T \rbrace$ can be approximated as follows:

Prior to observing $y_t$, the predictive distributions at time $t-1$ are
\begin{align*}
          Y_t \mid Y_{1:t-1}, \bPhi & \sim \Np{  f_t}{  Q_t} \\
    \btheta_t \mid Y_{1:t-1}, \bPhi & \sim \Np{\ba_t}{\bR_t} 
  \end{align*}
with
\begin{align*}
  \ba_t & = g ( \bm_{t-1}, \bzero )                        &
  \bR_t & = \bG_t \bC_{t-1} \bG_t^\prime + \bH_t \bW_t \bH_t^\prime  \\
    f_t & = f ( \ba_t, 0) &
    Q_t & = \bF_t^\prime \bR_t \bF_t     + \bJ_t V_t \bJ_t^\prime
\end{align*}
where
\begin{align*}
  \bG_t & = \frac{\partial g}{\partial \btheta} 
            \Bigr\rvert_{\hat{\btheta}_{t-1},\hat{\bw}_t}  &
  \bH_t & = \frac{\partial g}{\partial \bw} 
            \Bigr\rvert_{\hat{\btheta}_{t-1},\hat{\bw}_t}
\end{align*}
and
\begin{align*}
  \bF_t & = \frac{\partial f}{\partial \btheta} 
            \Bigr\rvert_{\hat{\btheta}_t,\hat{v}_t}  &
  \bJ_t & = \frac{\partial f}{\partial v} 
            \Bigr\rvert_{\hat{\btheta}_t,\hat{v}_t}.
\end{align*}
After observing $Y_t$, the posterior distribution of the state vector at time $t$ is
\begin{equation*}
  \btheta_t \mid Y_{1:t}, \bPhi \sim \Np{\bm_t}{\bC_t}
\end{equation*}
with
\begin{align*}
  \bm_t & = \ba_t + \bA_t e_t          &
  \bC_t & = \bR_t - \bA_t Q_t \bA_t^\prime
\end{align*}
where $e_t = Y_t - f_t$ and $\bA_t = \bR_t \bF_t / Q_t$.


\subsection*{Backward-sampling}

The joint posterior $\btheta_{1:T} \mid Y_{1:T}, \bPhi$ can be sampled recursively as follows:
\begin{itemize}
  \item Sample $\btheta_T^{(j)}$ from $\btheta_T \mid Y_{1:T}, \bPhi \sim \Np{\bm_T}{\bC_T}$
  \item for $k = 1,\ldots,T-1$
    \begin{itemize}
      \item Sample $\btheta_{T-k}^{(j)}$ from $\btheta_{T-k}^{(j)} \mid \btheta_{T-k+1}^{(j)}, Y_{1:T}, \bPhi \sim \Np{\bh_T(k)}{\bH_T(k)}$
    \end{itemize}
\end{itemize}
where
\begin{align*}
  \bh_T(k) 
    & = \bm_{T-k} + \bB_{T-k} \left( \btheta_{T-k+1}^{(j)} - \ba_{T-k+1} \right) \\
  \bH_T(k) 
    & = \bC_{T-k} - \bB_{T-k} \bR_{T-k+1} \bB_{T-k}^\prime
\end{align*}
and $\bB_{T-k} = \bC_{T-k} \bG_{T-k+1}^\prime \bR_{T-k+1}^{-1}$.
The quantities $\bm_t$, $\bC_t$, $\ba_t$, $\bR_t$ and $\bG_t$ are obtained from the filtering recursions.

\subsection*{Forecasting}

The $k$-step ahead forecast distribution given data up to time $t$ can be sampled sequentially as
\begin{itemize}
  \item Sample $\btheta_t^{(j)}$ from $\btheta_t \mid Y_{1:t}, M \sim \Np{\bm_t}{\bC_t}$
  \item for $i = 1,\ldots,k$
  \begin{itemize}
    \item Sample $\btheta_{t+i}^{(j)}$ from $g ( \btheta_{t+i-1}^{(j)}, \bw_{t+i} )$
    \item Sample $Y_{t+i}^{(j)}$ from $f ( \btheta_{t+i}^{(j)}, v_{t+i} )$.
  \end{itemize}
\end{itemize}

\bibliographystyle{plainnat}
\bibliography{library}
    
\end{document}


\maketitle

\section{Simulation study}

\begin{figure}[t]
  \includegraphics[width=0.49\textwidth]{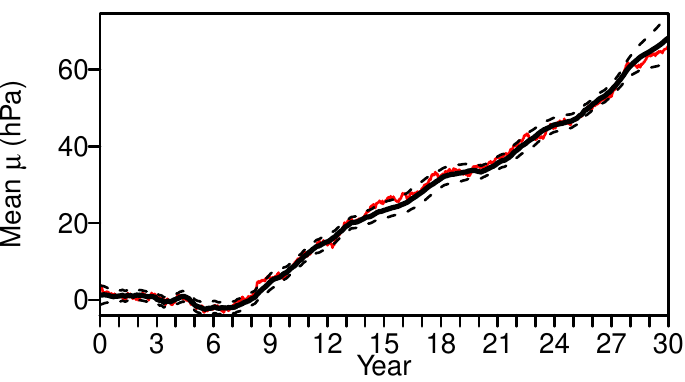}
  \includegraphics[width=0.49\textwidth]{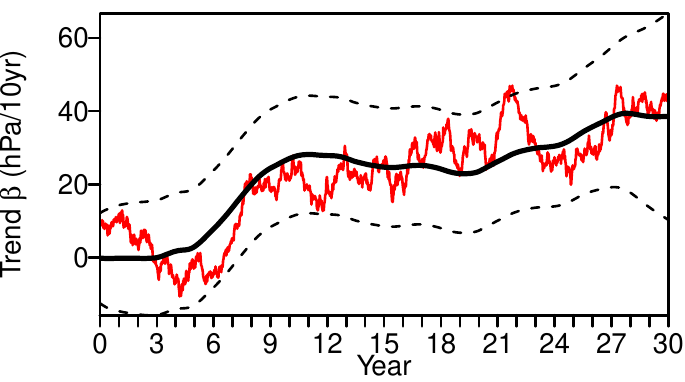} \\
  \includegraphics[width=0.49\textwidth]{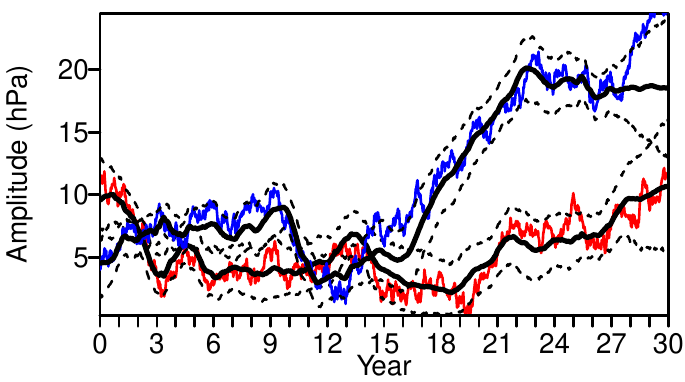}
  \includegraphics[width=0.49\textwidth]{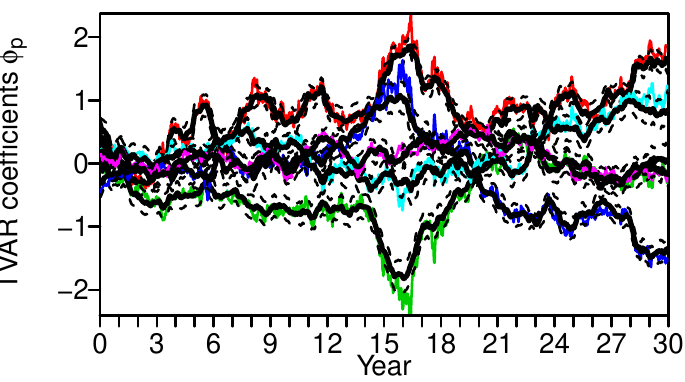}
  \caption{Simulation study under the null model.
           (top left) mean component $\mu_t$;
           (top right) trend component $\beta_t$;
           (bottom left) amplitude of seasonal components $\psi_1$ and $\psi_2$;
           (bottom right) TVAR coefficients $\phi_1,\ldots,\phi_5$.
           Solid black lines indicate the posterior mean, and dashed lines indicate a $95\,\%$ credible interval.
           Coloured lines indicate the true values.
           \label{fig:sims}}
\end{figure}

In order to test the linearised approximation proposed in the main text, we performed multiple simulation studies of the basic model defined by (1) and (2).
The performance in identifying the TVAR coefficients $\phi_1,\ldots,\phi_P$ is of particular interest.
To make the simulation study relevant to the NAO dataset analysed in the main text, we simulated daily data from a model with $K = 2$ harmonic components ($\omega = 2\pi/365$) and $P = 5$ TVAR coefficients.
To ensure that the autocorrelation structure remained stationary throughout, the TVAR coefficients were simulated from a time-varying partial-autocorrelation model 
\begin{align*}
 \rho_{p t} & = \rho_{p,t-1} + w_{\rho_p t} &
 w_{\rho_p t} & \sim \Np{0}{W_\rho} &
 p & = 1,\ldots,P
\end{align*}
with the constraint that $(-1 < \rho_{p t} < +1)$.
The time-varying partial-autocorrelation coefficients $\rho_{p t}$ are transformed into time-varying autoregressive coefficients using the Durbin-Levinson recursions \citep{Friedlander1982}.
A range of evolution variances were explored for the different model components, guided by the priors listed in Table~2 of the main text.
Figure~\ref{fig:sims} shows the results of a single simulation from the basic model defined by (1) and (2).
The prior probabilities for the state variables $\btheta_0$ are listed in Table~\ref{tab:sim-priors}, and were chosen to be only mildly informative.
The initial values of $\mu_0$, $\beta_0$, $\psi_{1,0}$, $\psi_{1,0}^\star$, $\psi_{2,0}$ and $\psi_{2,0}^\star$ were drawn from those prior distributions.
The evolution variances used are listed in Table~\ref{tab:sim-vars}, and were deliberately chosen to exceed the upper end of the prior ranges specified in Table~2 of the main text.
The evolution variances used for simulation were also used in the linearised filter.
In Figure~\ref{fig:sims}, the linearised filter is able to track all the model components, including the TVAR coefficients.
This performance is typical of that found during multiple simulation studies.
In repeated testing, the linear approximation proved itself to be remarkably robust for a wide range of evolution variances.
Full code is provided for further testing.

\begin{table}
  \caption{Prior probability distributions for state variables $\btheta_0$.
           All normally distributed.
           \label{tab:sim-priors}}
  \centering
  \begin{small}
    \begin{tabular}{lccc}
      \hline
      Component & Parameter & Mean & Variance \\
      \hline
      Mean level          & $\mu_0$                        & $0$ & $5^2$ \\
      Local trend         & $\beta_0$                      & $0$ & $0.002^2$ \\             
      Cyclic components   & $\psi_{1,0},\psi_{1,0}^\star,\psi_{2,0},\psi_{2,0}^\star$, & $0$ & $5^2$ \\
      Irregular component & $X_{-4},\ldots,X_0$            & $0$ & $10^2$ \\
      TVAR coefficients   & $\phi_{1,0},\ldots,\phi_{5,0}$ & $0$ & $1^2$ \\
      \hline
    \end{tabular}
  \end{small}
\end{table}

\begin{table}
  \caption{Evolution variances used in example simulation study.
           \label{tab:sim-vars}}
  \centering
  \begin{small}
    \begin{tabular}{lccc}
      \hline
      Component & Parameter & Variance & Log Variance \\
      \hline
      Observation variance & $V$       & $0.1^2$    & $- 4.6$ \\
      Mean variance        & $W_\mu$   & $0.1^2$    & $- 4.6$ \\     
      Trend variance       & $W_\beta$ & $0.0001^2$ & $-18.4$ \\
      Seasonal variance    & $W_\psi$  & $0.1^2$    & $- 4.6$ \\
      Irregular variance   & $W_X$     & $5.0^2$    & $+ 3.2$ \\
      Coefficient variance & $W_\phi$  & $0.015^2$  & $- 8.4$ \\
      \hline
    \end{tabular}
  \end{small}
\end{table}

\begin{figure}[t]
  \includegraphics[width=0.49\textwidth]{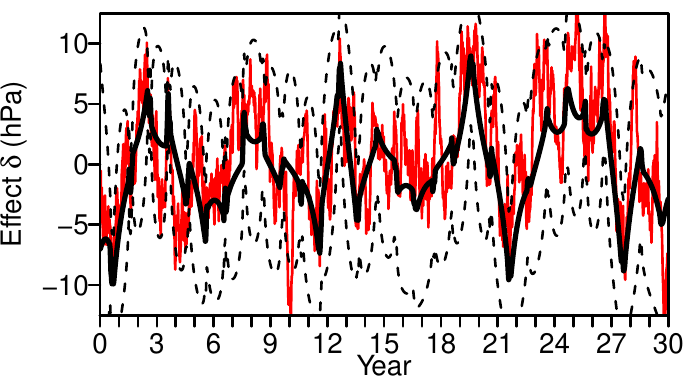}
  \includegraphics[width=0.49\textwidth]{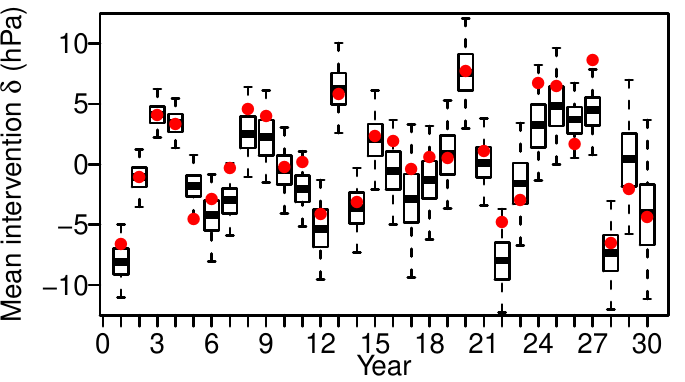}
  \caption{Simulation study under the mean intervention model $M_\mu$.
           (left) The intervention effect $\delta_t$.
           The solid black line indicates the posterior mean, and dashed lines indicate a $95\,\%$ credible interval.
           The red line indicates the true value.
           (right) The mean intervention effect.
           Boxes indicate the median and interquartile range inferred from the linearised filter.
           Whiskers indicate a $95\,\%$ credible interval.
           Red points indicate the true values.
           \label{fig:sims-mean}}
\end{figure}

We also tested the performance of the linearised filter for the mean intervention model $M_\mu$, defined by (1), (2), (4) and (5) in the main text.
Figure~\ref{fig:sims-mean} illustrates the performance for identifying the intervention effect $\delta_t$.
The same priors, evolution variances and random seed were used as in Figure~\ref{fig:sims}, so the mean, trend, seasonal and irregular and TVAR components are identical, and performance in identifying them is similar (not shown).
Figure~\ref{fig:sims-mean} includes an intervention lasting $\gamma = 90$ days with tapering factor $\rho = 0.2$, variance $W_\delta = 0.5^2$ and autocorrelation coefficient $\varphi = 0.995$.
The left hand plot showing the inferred evolution of the intervention effect $\delta_t$ is sharply spiked, the spikes indicating the $90$ days during which the forced effect influences the observations.
The right hand plot shows inference for the mean of the effect $\delta_t$ over the intervention period, similar to the main text.
The linearised filter shows excellent performance for identifying the intervention effect, the mean usually lying within the posterior inter-quartile range, and only once outside of the $95\,\%$ credible interval, as expected for a sample of $30$~years.
This performance is typical of that observed during testing.

\begin{figure}[t]
  \includegraphics[width=0.49\textwidth]{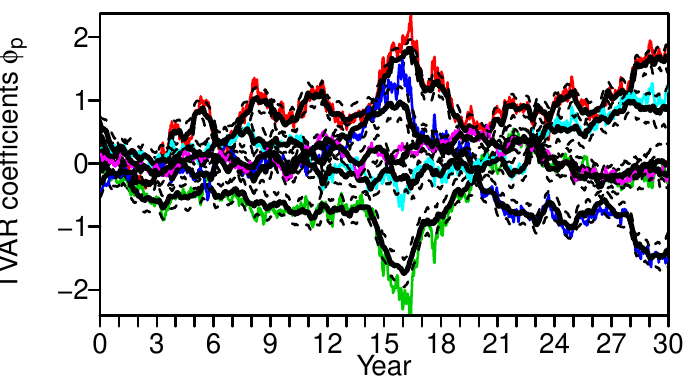}
  \includegraphics[width=0.49\textwidth]{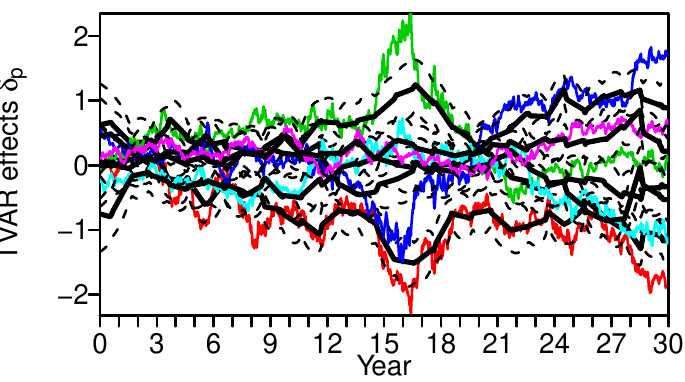} \\
  \includegraphics[width=0.49\textwidth]{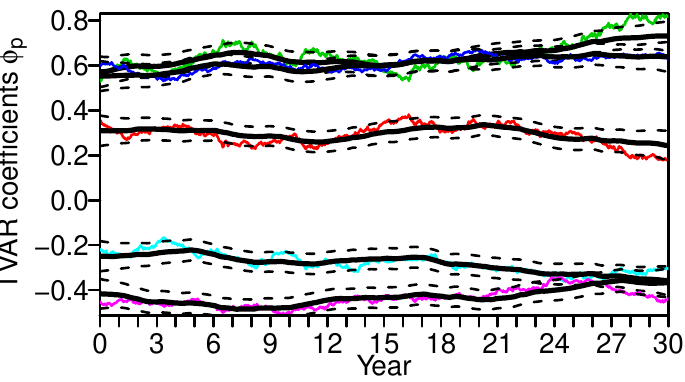}
  \includegraphics[width=0.49\textwidth]{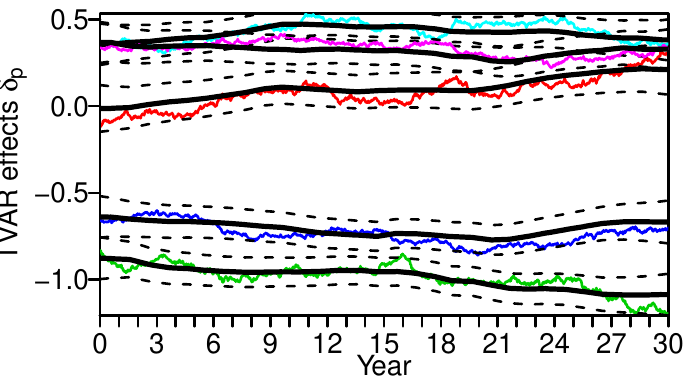}
  \caption{Simulation study under the autocorrelation intervention model $M_X$.
           (top) TVAR coefficients $\phi_1,\ldots,\phi_5$ and intervention effects $\delta_1,\ldots,\delta_5$ in a quickly evolving simulation $W_\phi \approx W_\delta \approx 0.015^2$;
           (bottom) TVAR coefficients $\phi_1,\ldots,\phi_5$ and intervention effects $\delta_1,\ldots,\delta_5$ in a slowly evolving simulation $W_\phi \approx W_\delta \approx 0.0015^2$
           Solid black lines indicate the posterior mean, and dashed lines indicate a $95\,\%$ credible interval.
           Coloured lines indicate the true values.
           \label{fig:sims-tvar}}
\end{figure}

The performance of the linearised filter for the autocorrelation intervention model $M_X$ defined by (1), (2), (6) and (7) was also tested.
Figure~\ref{fig:sims-mean} illustrates the performance for distinguishing the TVAR coefficients $\phi_{1 t},\ldots,\phi_{5 t}$ and intervention effects $\delta_{1 t},\ldots,\delta_{5 t}$.
The same priors, evolution variances and random seed were used as in Figures~\ref{fig:sims} and \ref{fig:sims-mean}, so the mean, trend and seasonal components are identical, and performance in identifying them is similar (not shown).
The top row of Figure~\ref{fig:sims-tvar} includes an intervention lasting $\gamma = 90$ days with tapering factor $\rho = 0.2$, variance $W_\delta = 0.001^2$ equal to $W_\phi$, and coefficient $\varphi = 1$.
The TVAR coefficients $\phi_{1 t},\ldots,\phi_{5 t}$ are identified well, but the linearised filter struggles with the intervention effects $\delta_{1 t},\ldots,\delta_{5 t}$.
In many simulations where the TVAR coefficients and the intervention effects are highly variable, the linearised filter struggled to track the intervention effects, failing to identify them at all or losing track at some point.
In the case of the NAO, and many other climate variables, we expect the TVAR coefficients to evolve more slowly, and any autocorrelation intervention effects to be similar every year.
The lower row of Figure~\ref{fig:sims-tvar} shows illustrates a simulation with with more slowly evolving TVAR and intervention coefficients.
The linearised filter is able to track both sets of coefficients accurately.
This is typical of the performance observed for slowly evolving cases, and gives us confidence that while the linearised filter has limitations, it performs well for the kind of situations we are interested in diagnosing.

\section{Markov chain Monte Carlo sampling}

A multivariate Normal proposal $q(\cdot \mid \cdot)$ was used for the hyper-parameters $\bPhi$, with initial diagonal covariance $\Sigma_0$ chosen by hand.
To encourage efficient mixing and ensure that all variance parameters were strictly positive, sampling was performed on the log of the variance parameters $V$, $W_\mu$, $W_\beta$, $W_\phi$, $c$ and $W_\delta$.
Sampling took place on the logit of the tapering parameter $\rho$ and coupled effect coefficient $\varphi$.
A Jacobian term was included in the computation of the acceptance probability to account for the transformation of $\varphi$ and $\rho$ relative to their priors.

Four chains were initialised from values chosen at random from the prior distribution of the hyper-parameters $\bPhi$.
Gelman-Rubin diagnostics were used to assess convergence of the chains and the effective sample size \citep{Gelman1992}.
The sampler was run in blocks of 1000 samples using the initial proposal distribution $\Sigma_0$ until all hyper-parameters achieved potential scale reduction factors of less than $2.0$.
Once approximate convergence was achieved, the sampler continued to run in blocks of 1000 samples, but using the adaptive Markov Chain Monte Carlo algorithm of \citep{Haario2001} to tune the proposal distribution until all hyper-parameters achieved potential scale factors of less than $1.1$.
Once convergence was achieved, the proposal distribution was fixed and the sampler continued to run in blocks of 1000 samples until the effective sample size exceeded 1000 samples for all hyper-parameters.
All previous samples were discarded.
For each chain, the mean acceptance rate after adaptation was around $0.30$.
To limit the computation, memory and storage requirements, backward sampling of the state parameters $\btheta_t$ ($t=1,\ldots,T$) was carried out separately for a subset of 1000 equally spaced values of $\bPhi$ from the converged sample set.

%

\begin{figure}[t]
  \includegraphics[width=\textwidth]{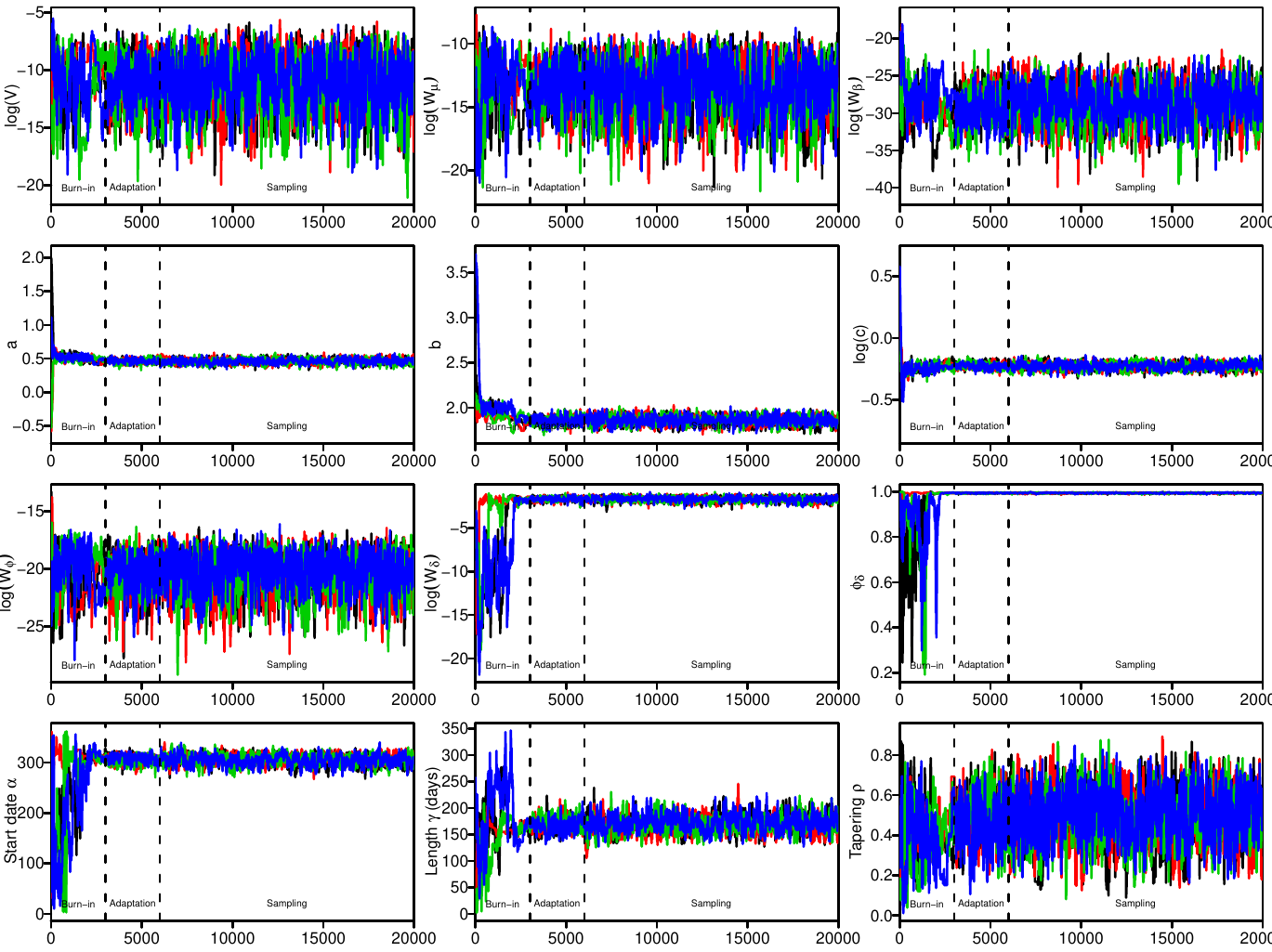}
  \caption{MCMC sample traces for the hyper-parameters $\bPhi$ from the mean intervention model $M_\mu$.
           Dashed lines indicate boundaries between samples from different chains.
           \label{fig:traces-mean}}  
\end{figure}

\begin{figure}[t]
  \includegraphics[width=\textwidth]{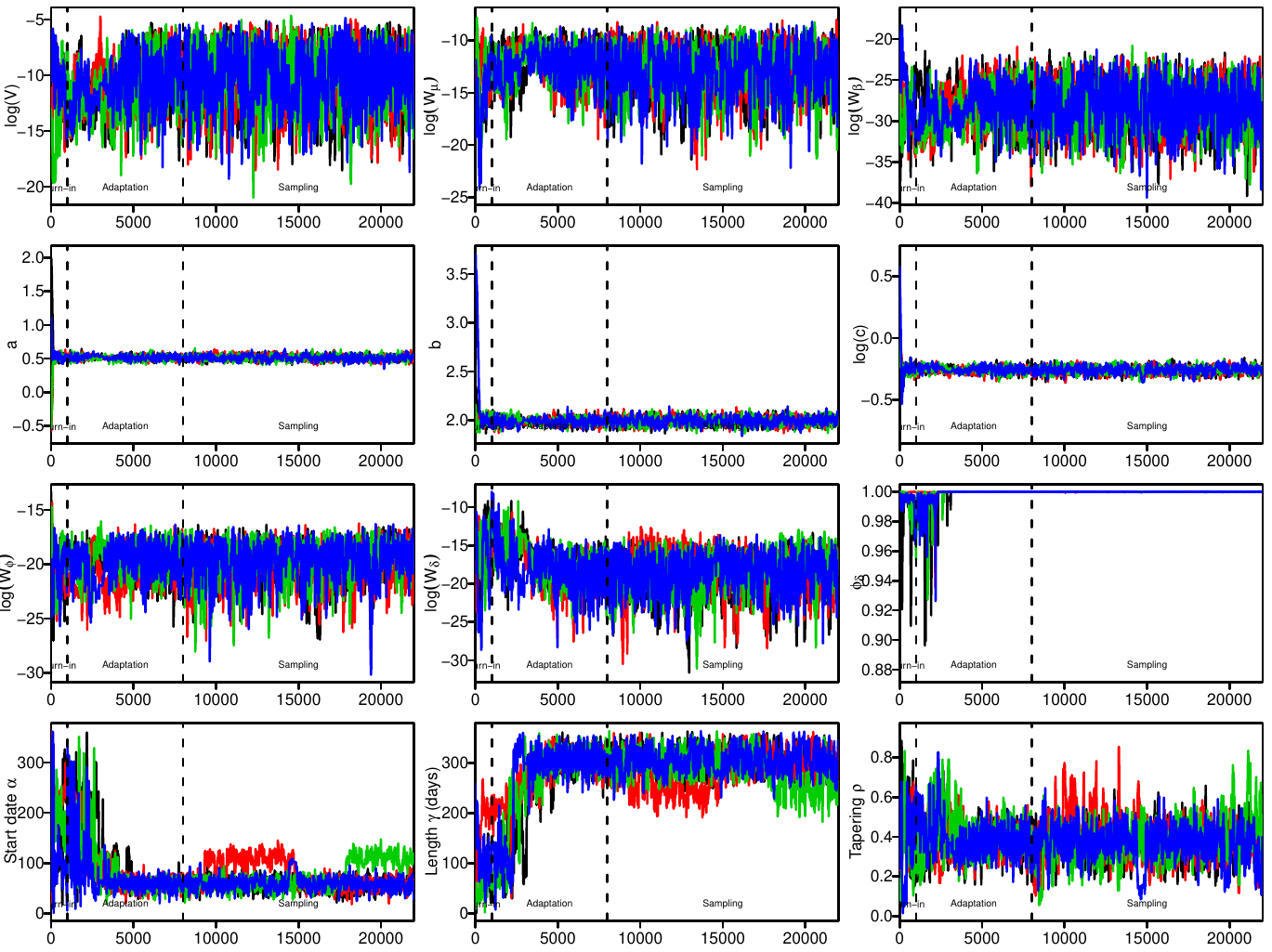}
  \caption{MCMC sample traces for the hyper-parameters $\bPhi$ from the autocorrelation intervention model $M_X$.
           Dashed lines indicate boundaries between samples from different chains.
           \label{fig:traces-x}}
\end{figure}

Trace plots of the posterior samples of each hyper-parameter from the mean intervention model are shown in Fig.~\ref{fig:traces-mean}.
All four chains converge to similar distributions and mix efficiently.
The traces from the autocorrelation intervention model in Fig.~\ref{fig:traces-x} behave similarly.

\begin{figure}[t]
  \includegraphics[width=\textwidth]{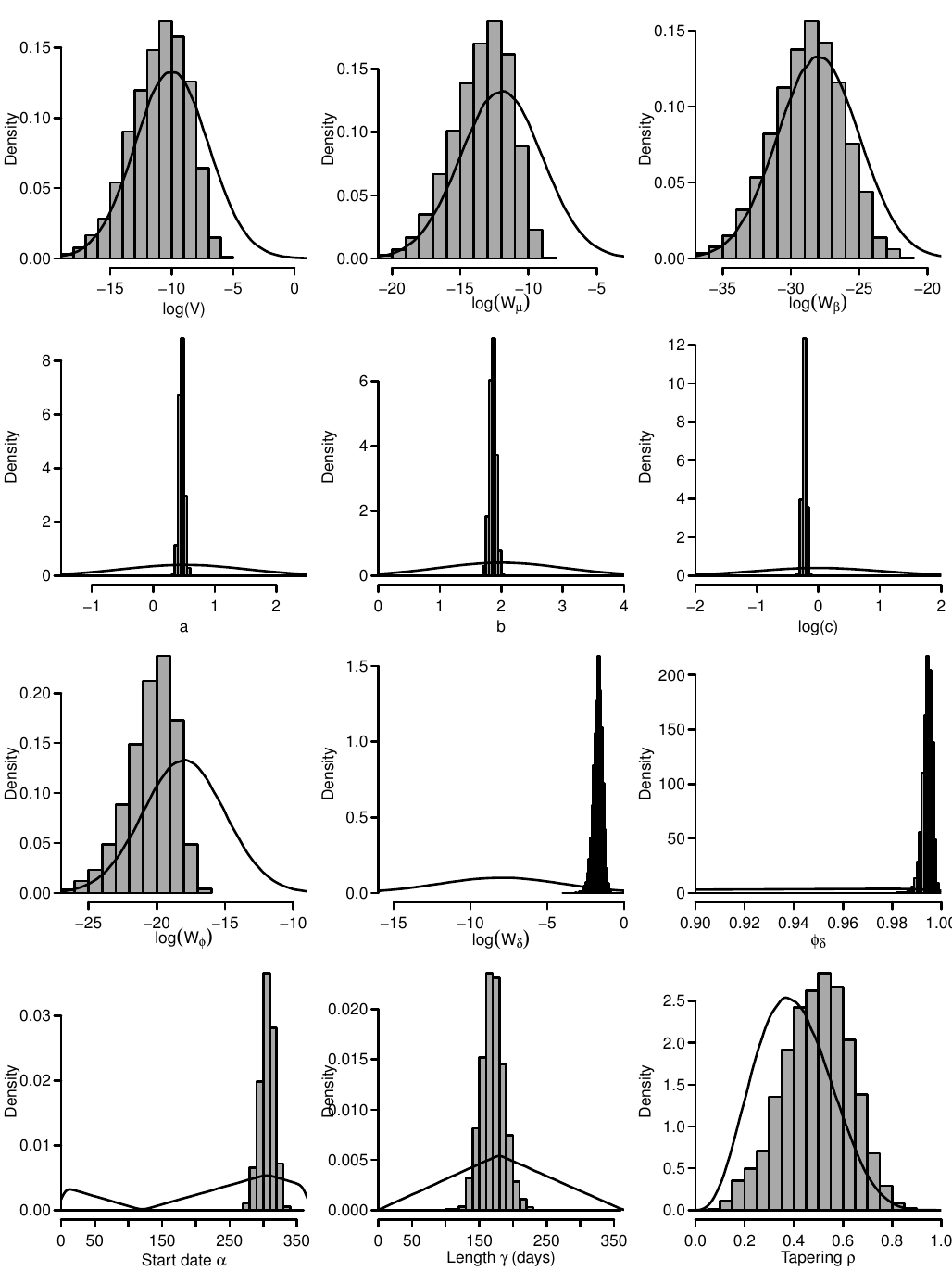}
  \caption{Posterior distributions of the hyper-parameters $\bPhi$ from the mean intervention model $M_\mu$.
           Grey histograms indicate the posterior densities.
           Black lines indicate the prior densities.
           \label{fig:hists-mean}}  
\end{figure}

\begin{figure}[t]
  \includegraphics[width=\textwidth]{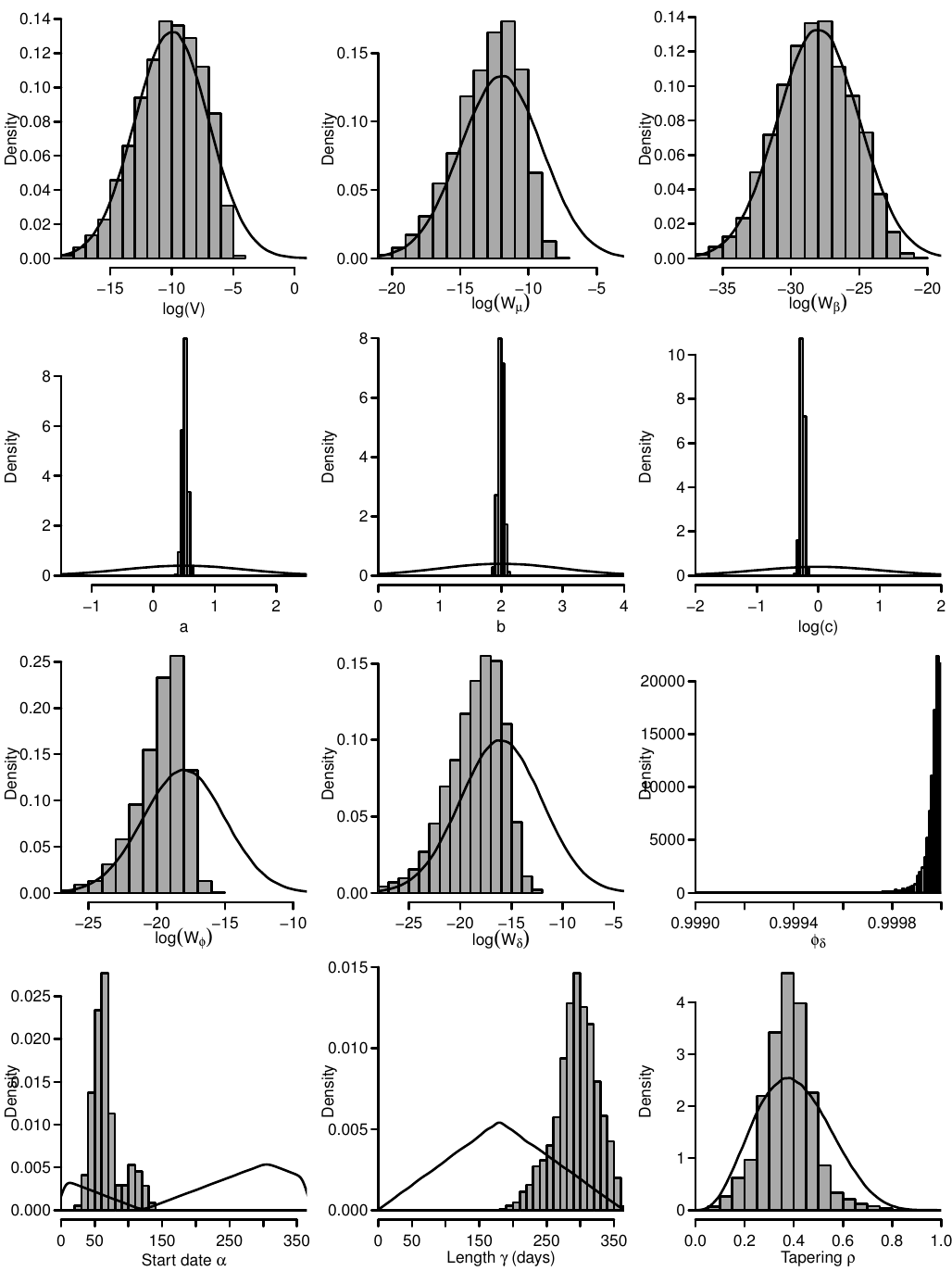}
  \caption{Posterior distributions of the hyper-parameters $\bPhi$ from the autocorrelation intervention model $M_X$.
           Grey histograms indicate the posterior densities.
           Black lines indicate the prior densities.
           \label{fig:hists-tvar}}
\end{figure}

Posterior density plots for the mean intervention model in Fig.~\ref{fig:hists-mean} indicate that the observations are very informative for the day-to-day variance parameters $a$, $b$ and $c$. 
The posterior distributions of the error variance $V$ and the innovation variances $W_\mu$ and $W_\beta$ somewhat reflect their respective prior distributions.
The posterior distributions do provide upper bounds on the error and innovation variances, putting useful limits on the amount of adaptation that might be expected from each component.
The posterior of $W_\phi$ is slightly more informative, and provides an upper bound on the amount of adaptation that might be expected from the AR parameters $\phi_{1 t},\ldots,\phi_{5 t}$.
The posterior samples from the autocorrelation intervention in Fig.~\ref{fig:hists-tvar} are broadly similar, although there is some evidence of bimodality in the intervention start date and length.

Neither model is able to usefully resolve the tapering parameter $\rho$, which broadly follows its prior distribution.
In the autocorrelation intervention model, the coefficient $\varphi$ is essentially unity and the effect variance $W_\delta$ is negligible, indicating a constant effect, similar for all intervention periods.
In the mean model the variance $W_\delta$ is well resolved and the coefficient $\varphi$ favour values slightly less that unity, indicating a volatile process with limited memory.

\section{Prior sensitivity}

\begin{figure}[t]
  \includegraphics[width=\textwidth]{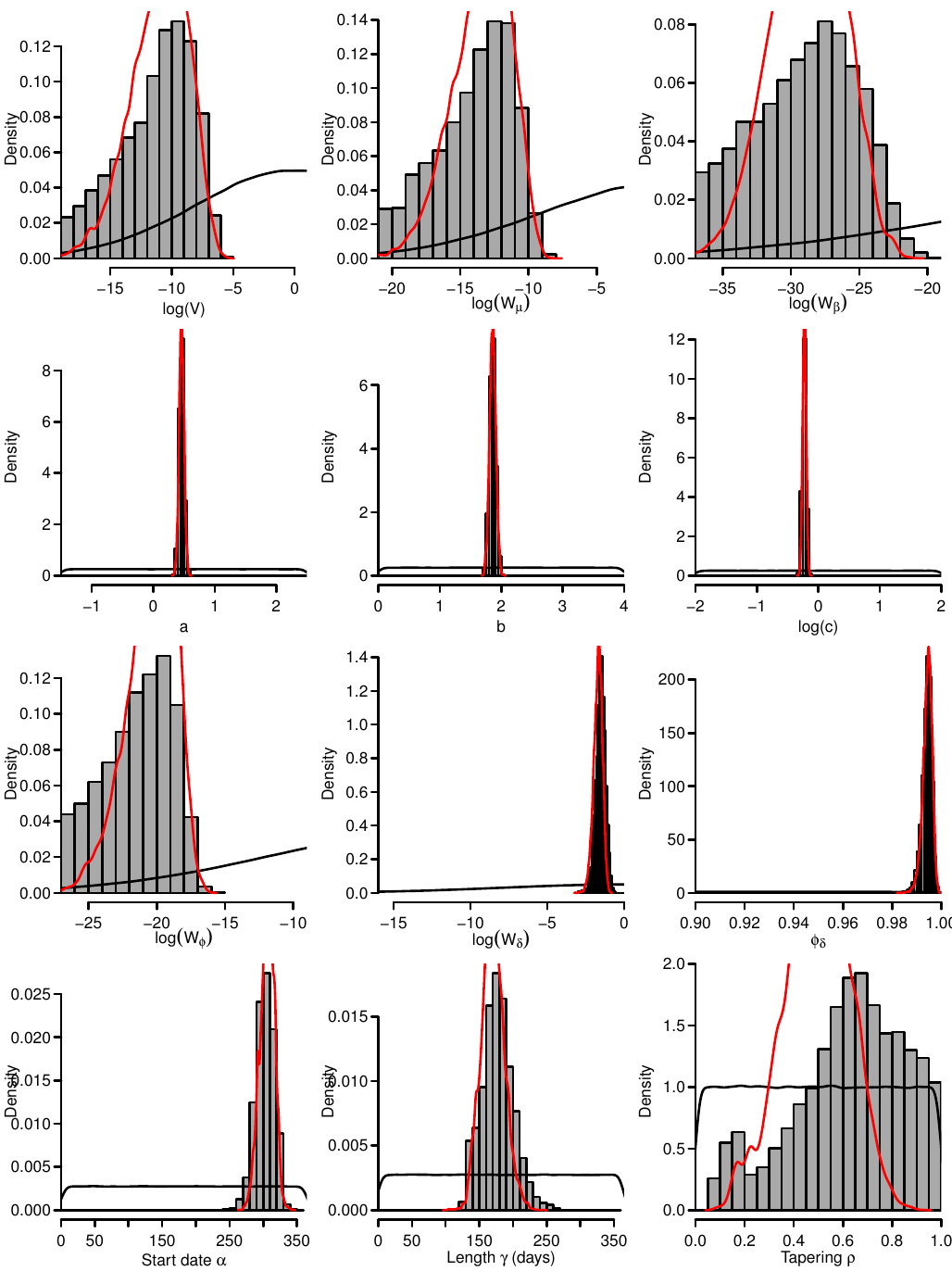}
  \caption{Posterior distributions of the hyper-parameters $\bPhi$ from the mean intervention model $M_\mu$ with the alternative priors.
           Grey histograms indicate the posterior densities.
           Black lines indicate the prior densities.
           Red lines indicate the posterior densities under the original priors.
           \label{fig:hists-mean-alt}}
\end{figure}

\begin{figure}[t]
  \includegraphics[width=\textwidth]{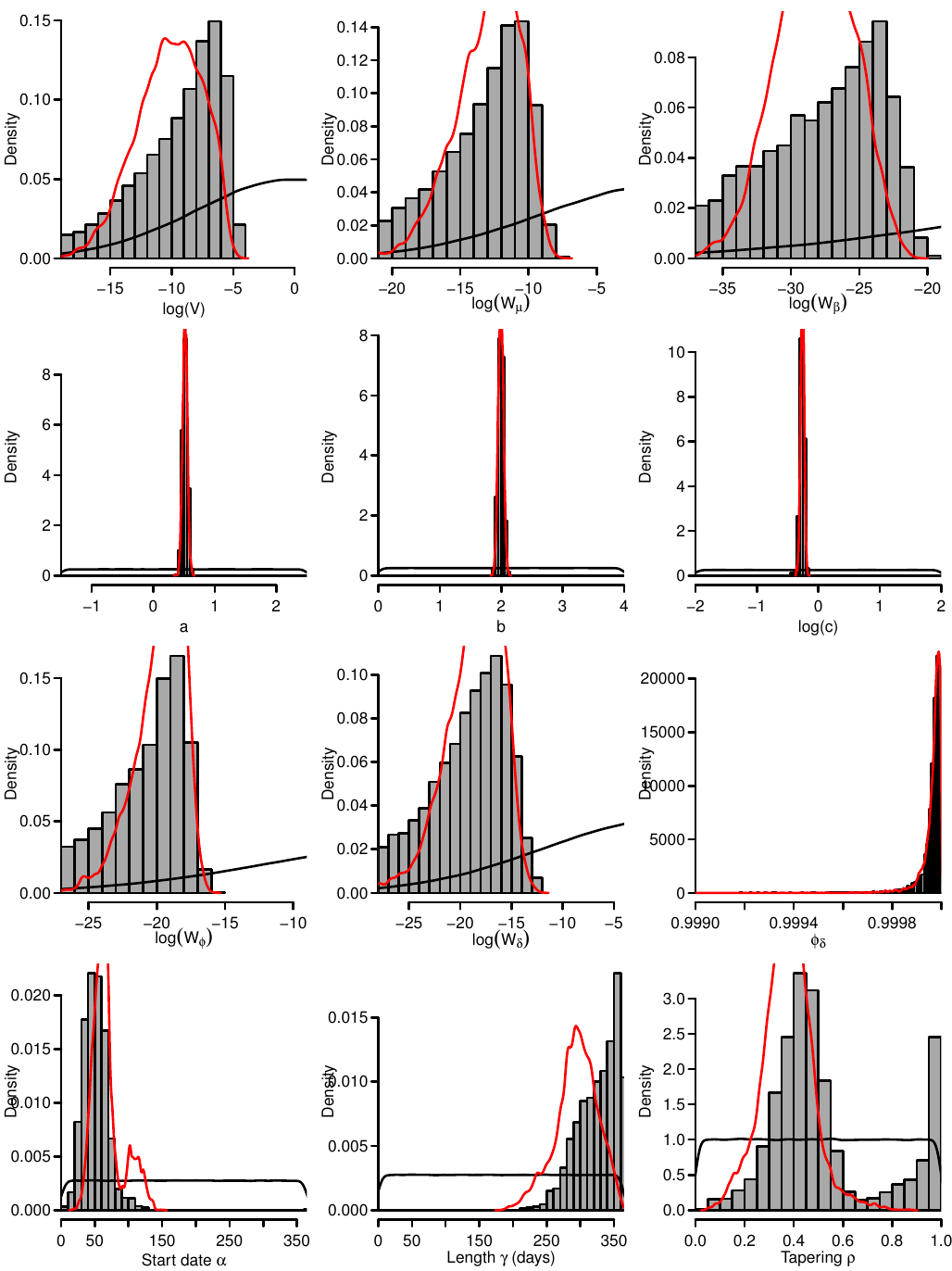}
  \caption{Posterior distributions of the hyper-parameters $\bPhi$ from the autocorrelation intervention model $M_X$ with the alternative priors.
           Grey histograms indicate the posterior densities.
           Black lines indicate the prior densities.
           Red lines indicate the posterior densities under the original priors.
           \label{fig:hists-tvar-alt}}
\end{figure}

To assess the sensitivity of our inferences to the priors on the hyper-parameters $\bPhi$, additional sampling runs were performed with flat priors for $a$, $b$, $\log W_X$, $\alpha$, $\gamma$, $\rho$, $\varphi_\mu$ and $\varphi_X$ and vague normal priors on the log-variances $\log V$, $\log W_\mu$, $\log W_\beta$, $\log W_\phi$, $\log W_{\delta_\mu}$ and $\log W_{\delta_\phi}$.
The alternative priors are listed in Tables~\ref{tab:variances} and \ref{tab:intervention}.
Figure~\ref{fig:hists-mean-alt} compares the posterior distribution of the hyper-parameters under the new priors and the original priors for the mean intervention model $M_\mu$.
The parameters $a$, $b$, $\log W_X$, $\log W_{\delta_\mu}$, $\varphi_\mu$, $\alpha$, $\gamma$ are all strongly constrained by the data in Fig.~\ref{fig:hists-mean}, and so their posterior distributions are almost identical under the weaker priors in Fig.~\ref{fig:hists-mean-alt}.
The tapering parameters $\rho$ is less well constrained by the data, and so its distribution is flattened slightly by the weaker prior.
Only the upper bounds of the variances log-variances $\log V$, $\log W_\mu$, $\log W_\beta$ and $\log W_\phi$ are strongly constrained by the data in.
Therefore, the lower tails of each posterior distribution lengthened to reflect the less informative prior distributions, but the upper tails remain basically unchanged.
The posterior distribution of the hyper-parameters $\bPhi$ under the autocorrelation intervention model $M_X$ are similarly unaffected by the vague priors in Fig.~\ref{fig:hists-tvar-alt}.

\begin{table}
  \caption{Prior densities for the hyper-parameters.
           \label{tab:variances}}
  \centering
  \begin{small}
    \begin{tabular}{lccc}
      \hline
      Component & Parameter & Prior & $\approx \SI{95}{\percent}$ Interval \\
      \hline
      Observation variance & $\log V$        & $\Np{0}{8^2}$ & $(-16,+16)$ \\
      Mean variance        & $\log W_\mu$    & $\Np{0}{9^2}$ & $(-18,+18)$ \\     
      Trend variance       & $\log W_\beta$  & $\Np{0}{17^2}$ & $(-34,+34)$ \\     
      Irregular variance   & $\log W_X$      & $\Up{-\infty}{+\infty}$  & $(-\infty,+\infty)$   \\
      Irregular variance   & $a$             & $\Up{-\infty}{+\infty}$ & $(-\infty,+\infty)$   \\
      Irregular variance   & $b$             & $\Up{-\infty}{+\infty}$ & $(-\infty,+\infty)$   \\     
      Coefficient variance & $\log W_\phi$   & $\Np{0}{12^2}$ & $(-24,+24)$ \\
      \hline
    \end{tabular}
  \end{small}
\end{table}

\begin{table}
  \caption{Prior densities for the intervention hyper-parameters.
          \label{tab:intervention}}
  \centering
  \begin{small}
    \begin{tabular}{lccc}
      \hline
      Component & Parameter & Prior & $\approx \SI{95}{\percent}$ Interval \\
      \hline
      Coupling start       & $\alpha$        & $\Up{0}{365}$   & $(0,365)$ \\
      Coupling length      & $\gamma$        & $\Up{0}{365}$   & $(0,365)$ \\
      Tapered proportion   & $\rho$          & $\Up{0}{1}$     & $(0.0,1.0)$   \\
      Mean effect variance & $\log W_{\delta_\mu}$ & $\Np{0}{8^2}$ & $(-16,+16)$ \\
      Mean effect coefficient   & $\varphi_\mu$       & $\Up{0}{1}$     & $(0.0,1.0)$   \\
      Autocorrelation effect variance & $\log W_{\delta_\phi}$ & $\Np{0}{12^2}$ & $(-24,+24)$ \\
      Autocorrelation effect coefficient   & $\varphi_\phi$       & $\Up{0}{1}$     & $(0.0,1.0)$   \\
      \hline
    \end{tabular}
  \end{small}
\end{table}

\begin{figure}[t]
  \centering
  \includegraphics[width=0.49\textwidth]{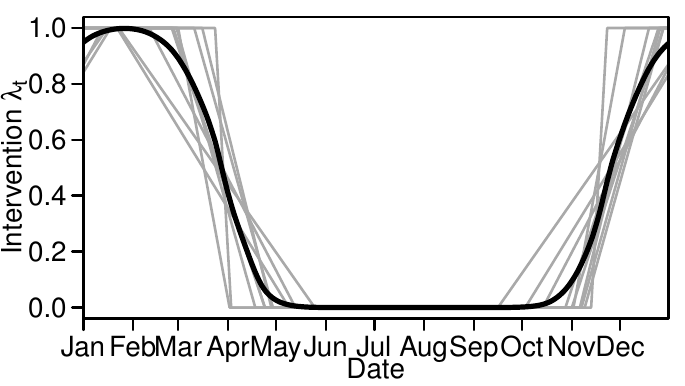}
  \includegraphics[width=0.49\textwidth]{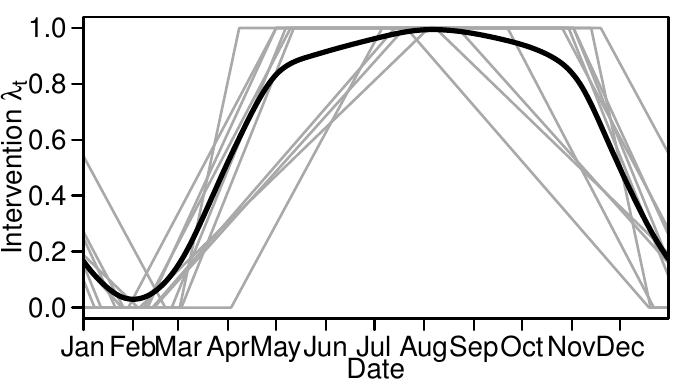}
  \caption{Posterior of the intervention $\lambda_t$ with the less informative priors.
           (left) The model with an intervention on the mean $M_\mu$;
           (right) the model with an intervention on the autocorrelation structure $M_X$.
           Grey lines represent a random sample of 10 realisations of the intervention $\lambda_t^{(j)}$ based on the posterior samples of $\alpha$, $\gamma$ and $\rho$.
           The black line is the pointwise posterior mean over all \num{1000} realisations of $\lambda_t^{(j)}$.
           \label{fig:vispost}}
\end{figure}

Figure~\ref{fig:vispost} shows that since the posterior distributions of the intervention start date $\alpha$ and length $\gamma$ are almost unchanged, the posterior distribution of the intervention $\lambda_t$ is also almost unchanged.

\begin{figure}[t]
  \includegraphics[width=0.49\textwidth]{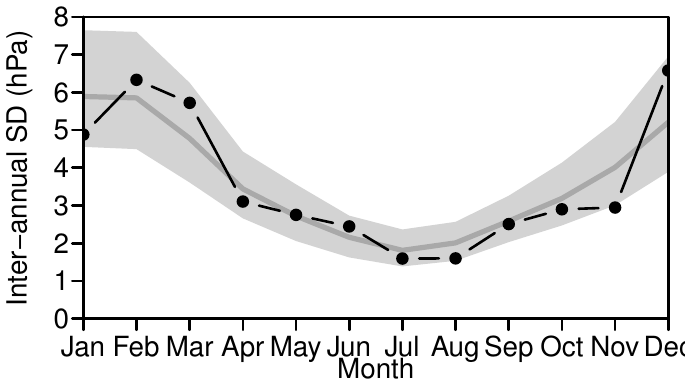}
  \includegraphics[width=0.49\textwidth]{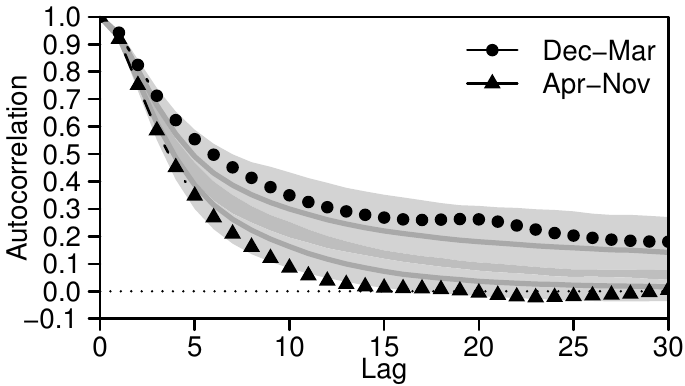} \\
  \includegraphics[width=0.49\textwidth]{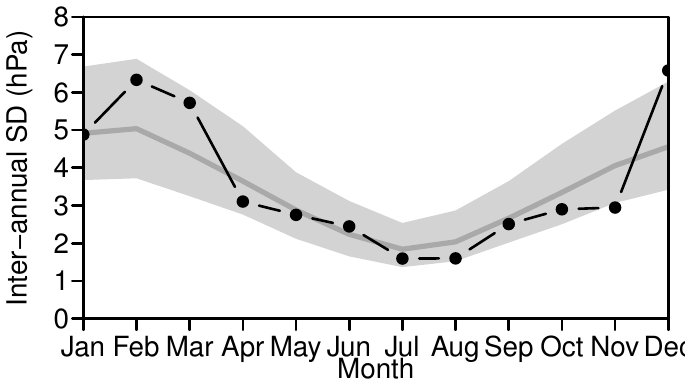}
  \includegraphics[width=0.49\textwidth]{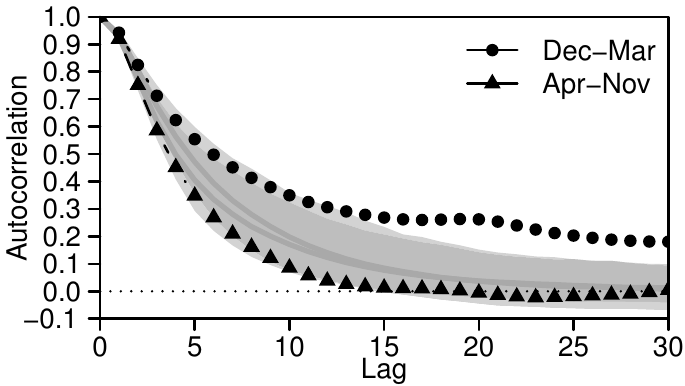}
  \caption{Posterior predictive checks with the less informative priors.
           (top) The model with an intervention on the mean $M_\mu$;
           (bottom) the model with an intervention on the autocorrelation structure $M_X$.
           Before computing the autocorrelation, the mean, a linear trend, annual and semi-annual cycles were estimated by least-squares and removed.
           Black lines represent the observed statistics.
           Dark grey lines indicate the posterior mean.
           Shading indicates pointwise \SI{90}{\percent} posterior credible intervals.
           Dark grey shading in (bottom right) indicates overlap between credible intervals.
           \label{fig:checks}}
\end{figure}

The posterior predictive checks  in Fig.~\ref{fig:checks} confirms that the mean intervention model $M_\mu$ is still able to reproduce the observed patterns of inter-annual variance and autocorrelation, but the autocorrelation intervention model $M_X$ cannot.
The Bayes' factor in favour of the mean intervention model $M_\mu$ increases to $B=37421$.

Since our priors on the log-variances $\log V$, $\log W_\mu$, $\log W_\beta$ and $\log W_\phi$ were chosen to have lower bounds corresponding to effectively zero variance, the inferences described in Section~6 of the main text were almost entirely unaffected by the changed priors.
The analysis of variance in Tab.~\ref{tab:anova} is essentially unchanged from that in the main text.

\begin{table}
  \caption{Analysis of variance. Bracketed values indicate \SI{90}{\percent} credible intervals.
           \label{tab:anova}}
  \centering
  \begin{small}
    \begin{tabular}{lcccc}
      \hline 
       & Mean & Coupling & Irregular & Error \\
      \hline
       Winter (Dec-Jan-Feb) & 0.00 (0.00,0.06) & 0.66 (0.52,0.77) 
                            & 0.33 (0.22,0.47) & 0.00 (0.00,0.00) \\
       Summer (Jun-Jul-Aug) & 0.15 (0.06,0.22) & 0.00 (0.00,0.00)
                            & 0.85 (0.78,0.94) & 0.00 (0.00,0.00) \\
      \hline
    \end{tabular}
  \end{small}
\end{table}

\begin{figure}[t]
  \includegraphics[width=0.49\textwidth]{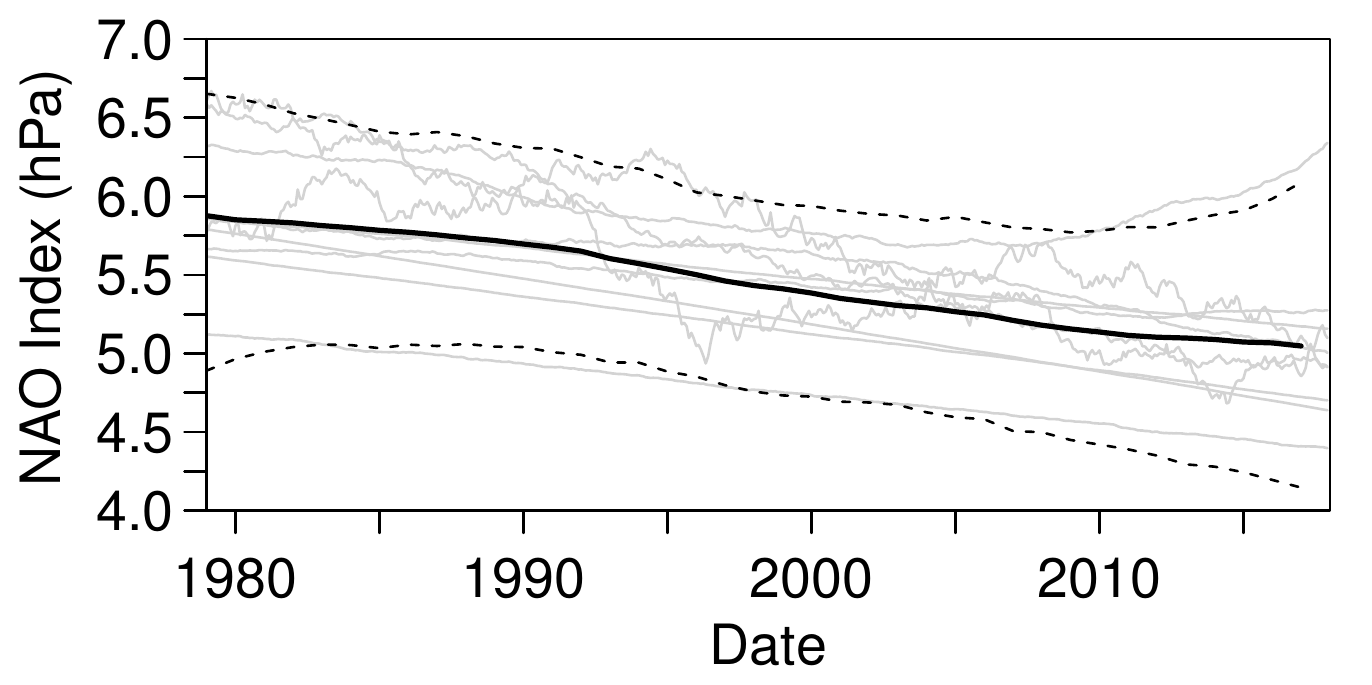}
  \includegraphics[width=0.49\textwidth]{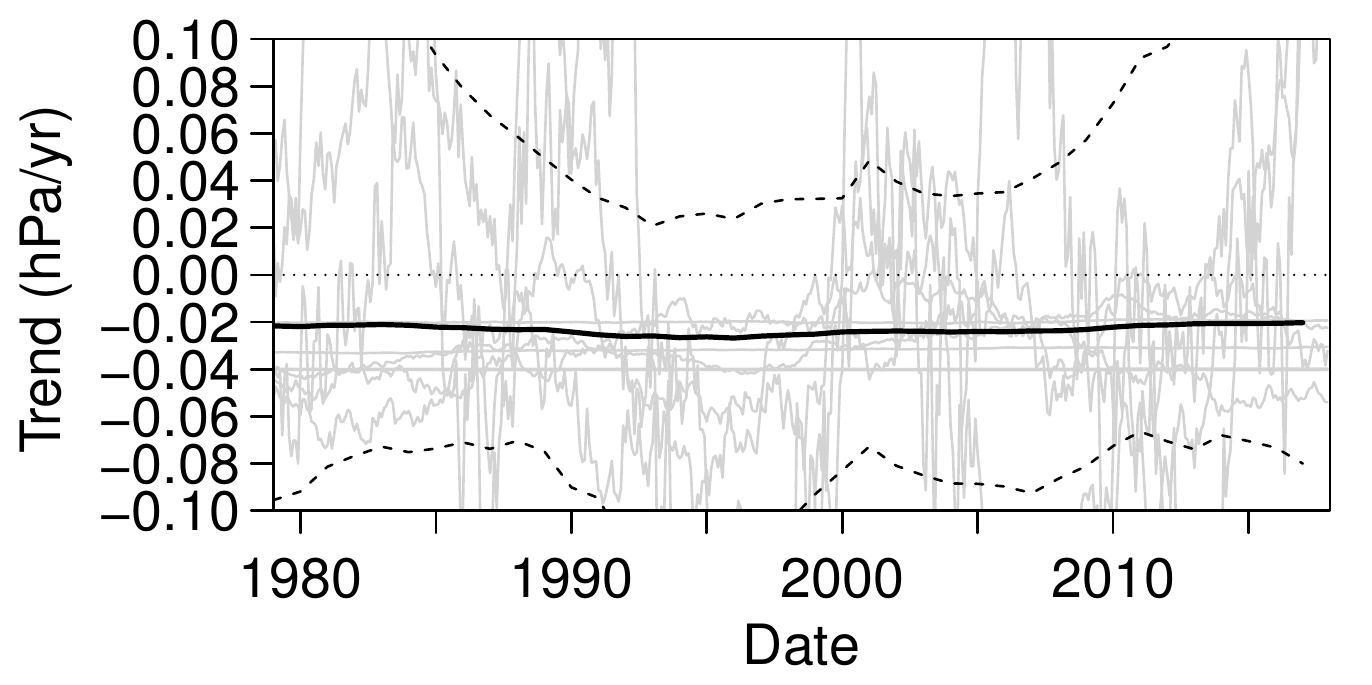}
  \caption{Posterior distributions of model components under the less informative priors.
           (left) Mean $\mu_t$;
           (right) trend $\beta_t$.
           Solid black lines represent the pointwise posterior mean.
           Dashed black lines represent pointwise \SI{90}{\percent} credible intervals.
           Grey lines are a random sample of 10 trajectories $\btheta_{1:T}^{(j)} \mid \bPhi^{(j)}, Y_{1:T}$.
           \label{fig:trends}}
\end{figure}

\begin{figure}[t]
  \includegraphics[width=\textwidth]{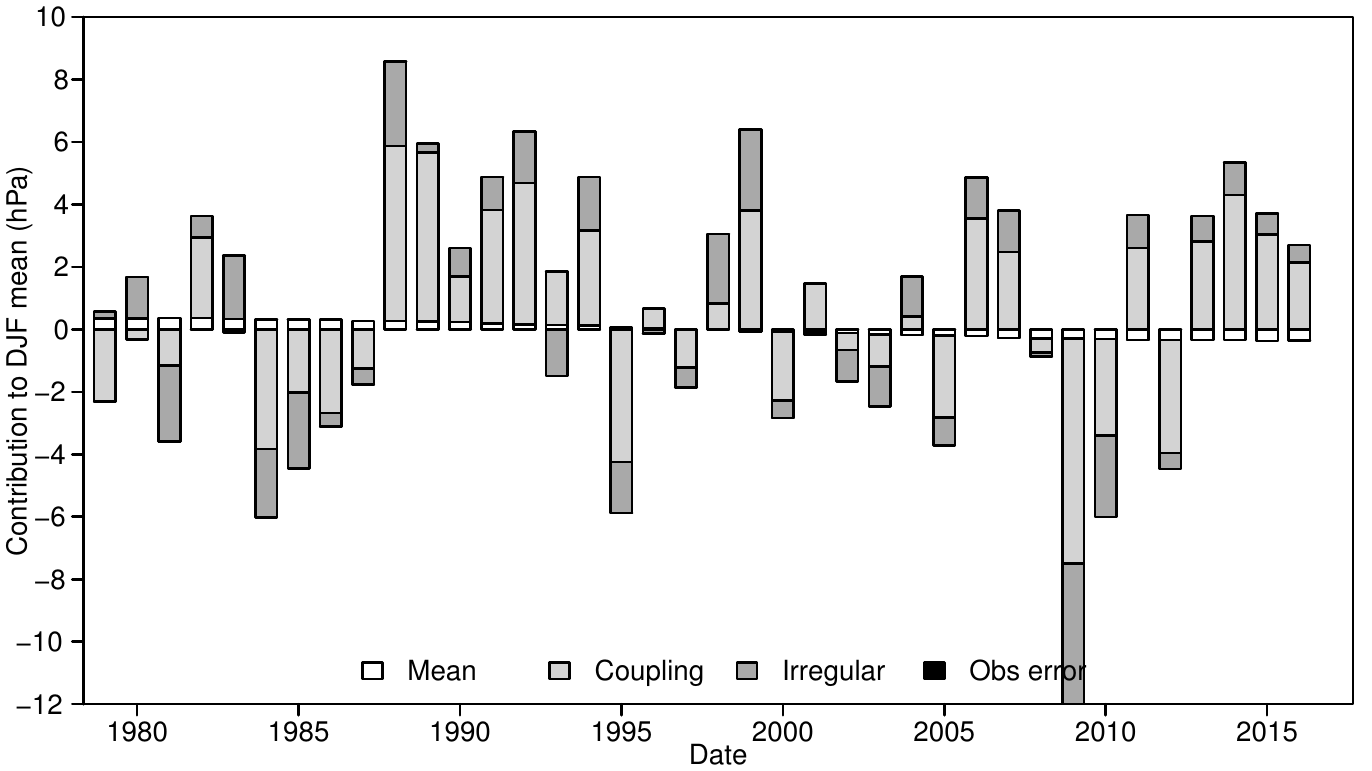}
  \caption{Contribution of individual model components under the less informative priors.
           Posterior mean estimates of the winter (Dec-Jan-Feb) mean levels of 
           the systematic component $\bar{\eta}$ ,
           the irregular component $\bar{X}$ ,
           the coupling effect $\bar{\delta}$, and
           the observation error $\bar{v}$.
           \label{fig:attrib}}
\end{figure}

The inferences for the posterior evolution of the mean $\mu_t$ in Fig.~\ref{fig:trends} and the contributions of the individual model components in Fig.~\ref{fig:attrib} are similarly unchanged.

\bibliographystyle{plainnat}
\bibliography{library}